\pdfoutput=1
\documentclass[fleqn,usenatbib,usedcolumn]{mnras}

\usepackage[british]{babel}  
\usepackage{amsmath,amssymb}   
\usepackage[T1]{fontenc}    
\usepackage{graphicx}  
   
\title[Stellar halo spin]{The slight spin of the old stellar halo}

\author[Deason et al.]{
Alis J. Deason$^{1}$\thanks{E-mail: alis.j.deason@durham.ac.uk}, Vasily Belokurov$^{2}$, Sergey E. Koposov$^{2,3}$, Facundo A. G\'{o}mez$^{4}$, 
\newauthor
\hspace{5pt}Robert J. Grand$^{5,6}$, Federico Marinacci$^{7}$, R\"{u}diger Pakmor$^5$\\
$^{1}$Institute for Computational Cosmology, Department of Physics, University of Durham, South Road, Durham DH1 3LE, UK\\
$^{2}$Institute of Astronomy, University of Cambridge, Madingley Road, Cambridge CB3 0HA, UK\\
$^{3}$McWilliams Center for Cosmology, Department of Physics, Carnegie Mellon University, 5000 Forbes Avenue, Pittsburgh, PA 15213, USA\\
$^{4}$Max-Planck-Institut f\"{u}r Astrophysik, Karl-Schwarzschild-Str. 1, D-85748, Garching, Germany  \\
$^{5}$ Heidelberger Institut f\"{u}r Theoretische Studien, Schloss-Wolfsbrunnenweg 35, 69118 Heidelberg, Germany\\
$^{6}$ Zentrum f\"{u}r Astronomie der Universit\"{a}t Heidelberg, Astronomisches Recheninstitut, M\"{o}nchhofstr. 12-14, 69120 Heidelberg, Germany\\
$^{7}$Department of Physics, Kavli Institute for Astrophysics and Space Research, MIT, Cambridge, MA 02139, USA}
\date{Accepted XXX. Received YYY; in original form ZZZ}

\pubyear{2017}

\begin{document}
\label{firstpage}
\pagerange{\pageref{firstpage}--\pageref{lastpage}}
\maketitle

\begin{abstract}

  We combine \textit{Gaia} data release 1 astrometry with Sloan Digital Sky Survey (SDSS) images taken some $\sim 10-15$ years earlier, to measure proper motions of stars in the halo of our Galaxy. The SDSS-\textit{Gaia} proper motions have typical statistical errors of 2 mas/yr down to $r \sim 20$ mag, and are robust to variations with magnitude and colour. Armed with this exquisite set of halo proper motions, we identify RR Lyrae, blue horizontal branch (BHB), and K giant stars in the halo, and measure their net rotation with respect to the Galactic disc. We find evidence for a gently rotating prograde signal ($\langle V_\phi \rangle \sim 5-25$ km s$^{-1}$) in the halo stars, which shows little variation with Galactocentric radius out to 50 kpc. The average rotation signal for the three populations is $\langle V_\phi \rangle = 14 \pm 2 \pm 10$ (syst.) km s$^{-1}$.  There is also tentative evidence for a kinematic correlation with metallicity, whereby the metal richer BHB and K giant stars have slightly stronger prograde rotation than the metal poorer stars. Using the Auriga simulation suite we find that the \textit{old} (T $>10$ Gyr) stars in the simulated halos exhibit mild prograde rotation, with little dependence on radius or metallicity, in general agreement with the observations. The weak halo rotation suggests that the Milky Way has a minor \textit{in situ} halo component, and has undergone a relatively quiet accretion history.

\end{abstract}


\section{Introduction}

Dark matter haloes have spin. This net angular momentum is acquired by tidal torquing in the early universe \citep{peebles69, dorosh70, white84}, and is later modified and shaped by the merging and accretion of substructures (e.g. \citealt{frenk85, catelan96, bullock01, gardner01, vitvitska02, peirani04, donghia07}). The acquisition and distribution of angular momenta in haloes is intimately linked to the evolution of the galaxies at their centres. Indeed, the relationship between halo spin and disc/baryonic spin is a fundamental topic in galaxy formation, and has been studied extensively in the literature (e.g. \citealt{vandenbosch02, sharma05, zavala08, bett10, deason11b, teklu15, zavala16}).

Initially, the angular momentum of the galaxy and the dark matter halo can be very well aligned. However, material is continually accreted onto the outer parts of the halo, which can alter its net angular momentum. Hence, while the galaxy and the halo often have aligned angular momentum vectors near their centers, they can be significantly misaligned at larger radii (e.g. \citealt{bett10, deason11b, gomez17}). Furthermore, major mergers can cause drastic ``spin flips'' in both the dark matter angular momenta and the central baryonic component \citep{bett12, padilla14}. 

It is clear that the net spin of haloes is critically linked to their merger histories, and thus their \textit{stellar haloes} could provide an important segue between the angular momenta of the central baryonic disc and the dark matter halo. A large fraction of the halo stars in our Galaxy are the tidal remnants of destroyed dwarfs. Hence, to first order, the spin of the Milky Way stellar halo represents the net angular momentum of all of its past (stellar) accretion events. 

The search for a rotation signal in the Milky Way halo dates back to the seminal work by \cite{frenk80}. The authors used line-of-sight velocities of the Galactic globular cluster system to infer a \textit{prograde} (i.e. aligned with the disc) rotation signal of  $V_{\rm rot} \sim 60$ km s$^{-1}$. A prograde signal, with $V_{\rm rot} \sim 40-60$ km s$^{-1}$,  in the (halo) globular cluster system has also been seen in several later studies (e.g. \citealt{zinn85, norris86, binney17}). However, the situation for the halo stars is far less clear. While most studies agree that the \textit{overall} rotation speed of the stellar halo is probably weak and close to zero \citep{gould98, sirko04, smith09, deason11a, fermani13b, das16}, there is some evidence for a kinematic correlation between metal-rich and metal-poor populations \citep{deason11a, kafle13, hattori13} and/or different rotation signals in the inner and outer halo \citep{carollo07,carollo10}.

An apparent kinematic dichotomy in the stellar halo (either inner vs. outer, or metal-rich vs. metal-poor) could be linked to different formation mechanisms. For example, state-of-the-art hydrodynamical simulations find that a significant fraction of the stellar haloes in the inner regions of Milky Way mass galaxies likely formed \textit{in situ}, and are more akin (at least kinematically) to a puffed up disc component \citep{zolotov09,font11,pillepich15}. Thus, one would expect a stronger prograde rotation signal in the inner and/or metal-rich regions of the Milky Way stellar halo \citep{mccarthy12}, and this theoretical scenario could account for the kinematic differences seen in the observations. However, as the detailed examination by \cite{fermani13b} shows, apparent kinematic signals depending on distance and/or metallicity can be wrongly inferred due to contamination in the halo star samples and/or systematic errors in the distance estimates to halo stars. Moreover, our observational inferences and comparisons with simulations should (but often do not) take into account the type of stars used to trace the halo. For example, commonly used tracers such as blue horizontal branch (BHB) and RR Lyrae (RRL) stars are biased towards old, metal-poor stellar populations, and this can affect the halo parameters we derive (see e.g. \citealt{xue11,janesh16}). 

So far, our examination of the kinematics of distant halo stars has been almost entirely based on one velocity component. For large enough samples over a wide area of sky, kinematic signatures such as rotation can be teased out using line-of-sight velocities alone. However, at larger and larger radii this line-of-sight component gives less and less information on the azimuthal velocities of the halo stars. Moreover, the presence of cold structures in line-of-sight velocity space \citep{schlaufman09} can also bias results. It is clearly more desirable to infer a direct rotation estimate from the 3D kinematics of the stars. Studies of distant halo stars with proper motion measurements are scarce \citep{deason13, koposov13, sohn15, sohn16}, but this limitation will become a distant memory as we enter the era of \textit{Gaia}. 

\textit{Gaia} is an unprecedented astrometric mission that will measure proper motions for hundreds of millions of stars in our Galaxy. In this contribution, we exploit the first data release of \textit{Gaia} (DR1, \citealt{gaia16}) to measure the net rotation of the Milky Way stellar halo. Although the first \textit{Gaia} data release does not contain any proper motions, we combine the exquisite astrometry of DR1 with the Sloan Digital Sky Survey (SDSS) images taken some $\sim 10-15$ years earlier to provide a stable and robust catalog of proper motions. Halo star tracers that have previously been identified in the literature are cross-matched with this new proper motion catalog to create a sample of halo stars with 2/3D kinematics.

The paper is arranged as follows. In Section \ref{sec:pms} we introduce the SDSS-\textit{Gaia} proper motion catalogue and investigate the statistical and systematic uncertainties in these measurements using spectroscopically confirmed QSOs. Our halo star samples are described in Section \ref{sec:samples}, and we provide further validation of our proper motion measurements by comparison with models and observations of the Sagittarius stream in Section \ref{sec:sgr}. In Section \ref{sec:like}, we introduce our rotating stellar halo model and apply a likelihood analysis to RRL, BHB and K giant halo star samples. We compare our results with state-of-the-art simulations in Section \ref{sec:sims}, and re-evaluate our expectations for the stellar halo spin. Finally, we summarise our main conclusions in Section \ref{sec:conc}.

\section{SDSS-\textit{Gaia} Proper Motions}
\label{sec:pms}

The aim of this work is to infer the average rotation signal of the Galactic halo using a newly calibrated SDSS-\textit{Gaia} catalog. This catalog (described below) is robust to systematic biases, which is vital in order to measure a rotation signal. Indeed, even with large proper motion errors (of order the size of the proper motions themselves!), with large enough samples distributed over the sky, the rotation signal can still be recovered provided that the errors are largely random rather than systematic.\\

The details of the creation of the recalibrated SDSS astrometric catalogue and measurement of SDSS-{\it Gaia} proper motions will be described in a separate paper (Koposov 2017 in preparation), but here we give a brief summary of the procedure. 

In the original calibration of the astrometry of SDSS sources, exposed in detail by \cite{pier03}, there are two key ingredients. The first is the mapping between pixel coordinates on the CCD $(x,y)$ and the coordinates corrected for the differential chromatic refraction and distortion of the camera $(x',y')$ (see Eqn. 5-10 in \citealt{pier03}). The second is the mapping between $(x',y')$ and the great circle coordinates on the sky $(\mu, \nu)$ aligned with the SDSS stripe (Eqn. 9, 10, 13, 14 of \citealt{pier03}). The first transformation does not change strongly with time, requires only a few free parameters and is well determined in SDSS. However, the second transformation that describes the scanning of the telescope, how non-uniform it is and how it deviates from a great circle, as well as the behaviour of anomalous refraction is much harder to measure. In fact, the anomalous refraction and its variation at small timescales is the most dominant effect limiting the quality of SDSS astrometry (see Fig. 13 of \citealt{pier03}). The reason why those systematic effects could not have been properly addressed by the SDSS project itself is that the density of astrometric standards from UCAC
\citep{zacharias13} and Tycho catalogues used for the derivation of the $(x',y')$, $(\mu,\nu)$ transformation was too low. This is where the \textit{Gaia} DR1 comes to the rescue, with its astrometric catalogue being $\sim$ 4 magnitudes deeper than UCAC. The only issue with using the \textit{Gaia} DR1 catalogue as a reference for SDSS calibration is that the epoch of the \textit{Gaia} catalogue is 2015.0 as opposed to $\sim$ 2005 for SDSS and that the proper motions are not yet available for the majority of \textit{Gaia} DR1 stars.

To address this issue, we first compute the relative proper-motions between \textit{Gaia} and the original SDSS positions in bins in color-magnitude space and pixels on the sky (HEALPix level
16, angular resolution 3.6 deg; \citealt{gorski05}) that gives us
estimates of $\langle \mu_{\alpha}( \mathrm{hpx}, g-i,i) ]\rangle$
  $\langle \mu_{\delta}(\mathrm{hpx},g-i,i) \rangle$.  Those average proper motions can be used to estimate the expected positions of \textit{Gaia} stars at the epoch of each SDSS scan.
  \begin{equation}
    \hat\alpha_{\rm SDSS} = \alpha_{Gaia} - \langle \mu_{\alpha}(\mathrm{hpx},g-i,i)  \rangle \delta T
    \end{equation}
  where $\delta T$ is the timespan between \textit{Gaia} and SDSS observation of a given star, hpx is the HEALPix pixel number of the star and $g-i$, and $i$ are colors and magnitudes of the star. With those positions $(\hat{\alpha}_{\rm SDSS}, \hat{\delta}_{\rm SDSS})$ computed for all the stars with both SDSS and \textit{Gaia} measurements we redetermine the astrometric mapping in SDSS between $(x',y')$ pixel coordinates and on the sky great circle $(\mu,\nu)$ coordinates by using a flexible spline model. There are many more stars available in \textit{Gaia} DR1 compared to the UCAC catalog, so in the model we are able to much better describe the anomalous refraction along the SDSS scans and, therefore, noticeably reduce the systematic uncertainties of the astrometric calibration. Furthermore, as a final step of the calibration, we also utilise the galaxies observed by Gaia and SDSS to remove any residual large scale astrometric offsets in the calibrated SDSS astrometry. With the SDSS astrometry recalibrated, the SDSS-{\it Gaia} proper motions are then simply obtained from the \textit{Gaia} positions and their recalibrated position in SDSS.

\subsection{Proper motion errors}
\label{sec:pmerr}
\begin{figure}
    \centering
    \includegraphics[width=8.5cm, height=7.08cm]{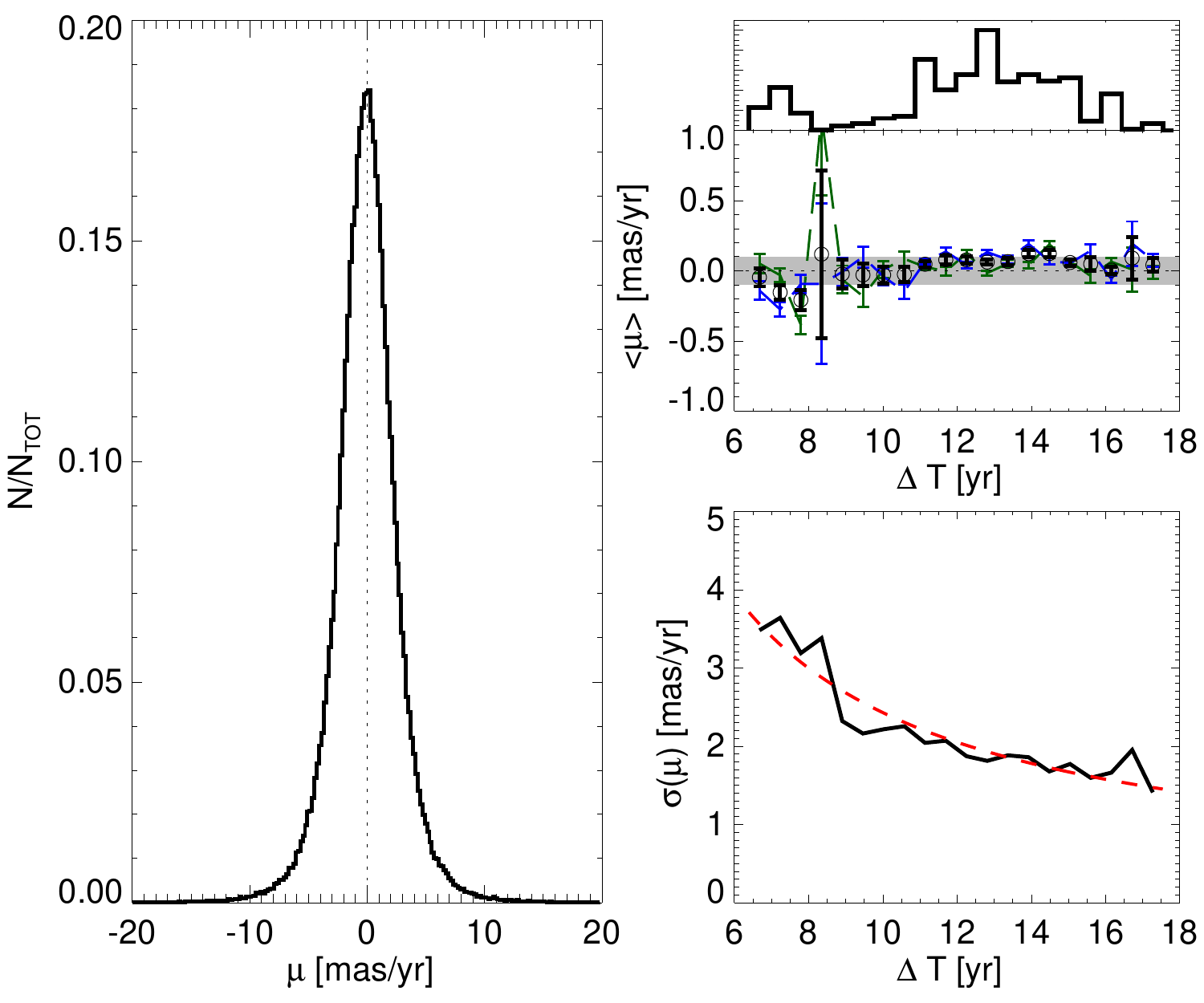}
    \caption[]{ \small \textit{Left panel:} The distribution of measured proper motions of SDSS DR12 spectroscopically confirmed QSOs. We find very similar distributions for $\mu_{\alpha}$ and $\mu_\delta$ (also $\mu_\ell$ and $\mu_b$), so for simplicity we use both proper motion measurements in this plot (i.e. $\mu=[\mu_\alpha, \mu_\delta]$). \textit{Top right panel:} A histogram of the time baseline between first epoch SDSS and second epoch \textit{Gaia} measurements ($\Delta T$). \textit{Middle right panel:} Median proper motion of QSOs as a function of time baseline. The median $\mu_{\alpha}$ and $\mu_\delta$ values are shown with the dashed green and blue lines respectively. The median proper motions are consistent with zero at the 0.1 mas/yr level. The grey shaded region indicates median offsets from zero of $\pm 0.1$ mas/yr.  \textit{Bottom right panel:} The dispersion in QSO proper motions as a function of $\Delta T$. Here, $\sigma$ is 1.48 times the median absolute deviation. The red dashed line shows the best-fit model for $\sigma(\mu)$, where $\sigma= A+B/\Delta T$. We use this relation to assign proper motion uncertainties to stars in the SDSS-\textit{Gaia} sample as a function of $\Delta T$.}
    \label{fig:qso_mjd}
\end{figure}

\begin{figure}
    \centering
    \includegraphics[width=8.5cm, height=4.25cm]{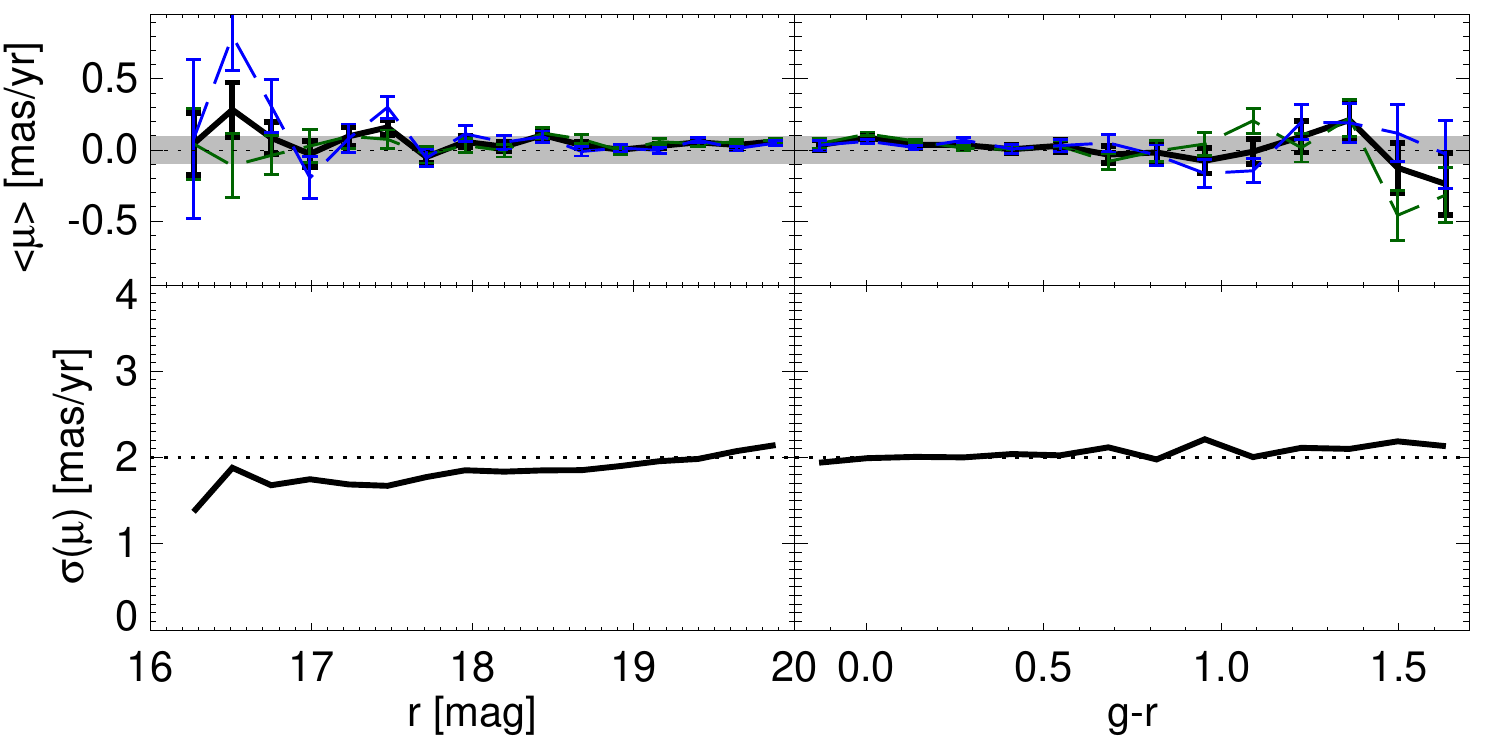}
    \caption{ \small Proper motion errors estimated from SDSS DR12 QSOs as a function of $r$-band magnitude (left panel) and $g-r$ colour (right panel). The top and bottom panels show the median and standard deviation of the QSO proper motions. The dashed green and blue lines in the top panels show the median $\mu_{\alpha}$ and $\mu_\delta$ proper motions, and the grey shaded region indicates median offsets from zero of $\pm 0.1$ mas/yr. The dotted line in the bottom panels indicates the median proper motion error of 2 mas/yr. There is a slight correlation of $\sigma(\mu)$with $r$-band magnitude, but this is very minor over the magnitude range probed in this study ($ r \lesssim 19$). Furthermore, there is no variation with colour. }
    \label{fig:qso_mag_col}
\end{figure}

We quantify the uncertainties in the SDSS-\textit{Gaia} proper motion measurements using spectroscopically confirmed QSOs from SDSS DR12 \citep{paris17}. This QSO sample is cross-matched with the SDSS-\textit{Gaia} catalog by searching for the nearest neighbour within $1\arcsec$. There are $N=71, 799$ QSOs in the catalog with $r < 20$, and we show the distribution of  QSO proper motions in the left-hand panel of Fig. \ref{fig:qso_mjd}. The QSO proper motions are nicely centred around $\mu =0$ mas/yr, and there are no significant high proper motion tails to the distribution. Note that we find no significant differences between the QSO proper motion components $\mu_\alpha$ and $\mu_\delta$, so we group both components together (i.e. $\mu=[\mu_\alpha, \mu_\delta]$) in the figure. However, we do show the $\mu_\alpha$ and $\mu_\delta$ components separately (green and blue dashed lines in the top-right panel) when we show the median proper motions to illustrate that these components \textit{individually} have no significant systematics.

The proper motion errors should roughly scale as $\sigma (\mu) \propto 1/\Delta T$, where $\Delta T$ is the timescale between the first epoch SDSS measurements and the second epoch \textit{Gaia} data\footnote{Note we compute $\Delta T$ using the modified Julian dates (MJD) of the SDSS observations and the last date of data collection for \textit{Gaia} DR1, i.e $\Delta T = $ MJD(\textit{Gaia})-MJD(SDSS) where MJD(\textit{Gaia})=MJD(16/9/2015)}. The SDSS photometry was taken over a significant period of time, and data from later releases have shorter time baselines. Thus, this variation in astrometry timespan is an important parameter when quantifying the proper motion uncertainties in our SDSS-\textit{Gaia} catalog. The top-right panel of Fig. \ref{fig:qso_mjd} shows a (normalised) histogram of the time baselines ($\Delta T$). There is a wide range of time baselines, but most of the SDSS data were taken $\sim 10-15$ years ago. In the bottom-right panel of Fig. \ref{fig:qso_mjd} we show the dispersion in QSO proper motion measurements (defined as $\sigma = 1.48$ times the median absolute deviation) as a function of $\Delta T$, and the middle-right panel shows the median values. The median values are consistent with zero at the level of $\sim 0.1$ mas/yr, and there is no systematic dependence on $\Delta T$.  As expected, there is a strong correlation between the dispersion of QSO proper motions and $\Delta T$. The dashed red line shows a model fit to the relation of the form: 

\begin{equation}
\label{eqn:sig}
\sigma = A + B/\Delta T,
\end{equation}

where $A=0.157$ mas/yr and $B=22.730$ mas. It is encouraging that this simple $A+B/\Delta T$  model agrees well with the QSO data, and we find no significant systematic differences between different SDSS data releases. Note that we show in Appendix \ref{sec:appendix} that there is no significant systematic variation in the QSO proper motions with position on the sky.

\begin{figure*}
    \centering
    \includegraphics[width=16cm, height=5.33cm]{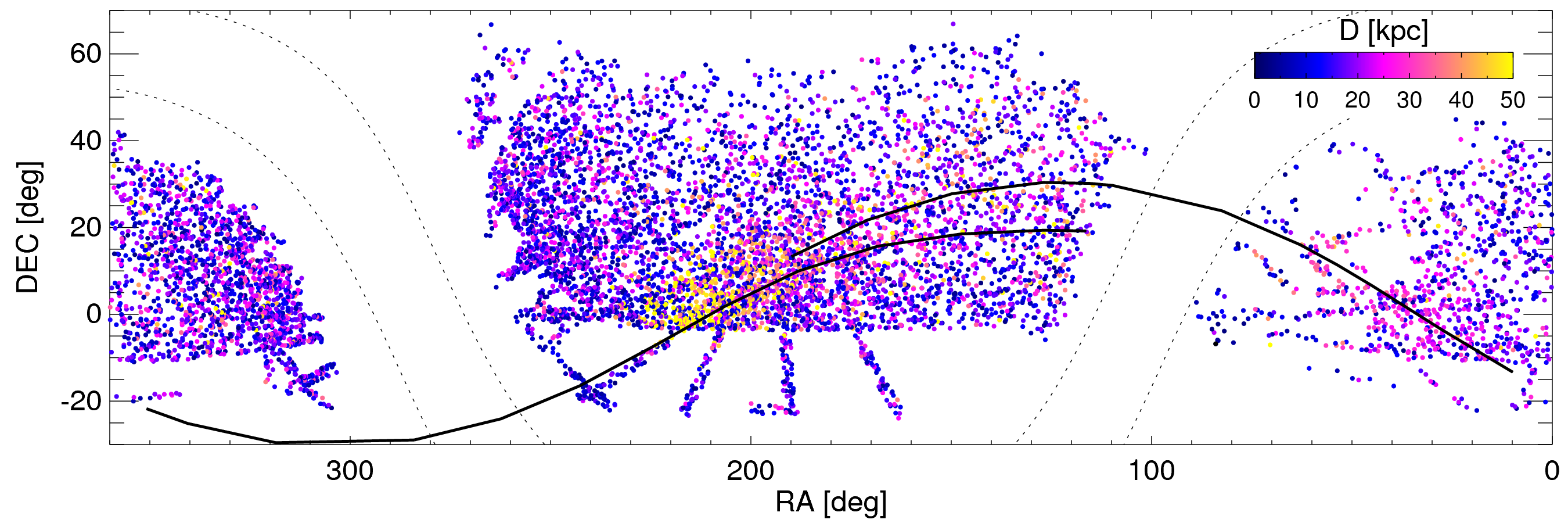}
    \caption[]{\small The sky distribution of ($N=8590$) RRL stars with SDSS-\textit{Gaia} proper motions in Equatorial coordinates. The points are colour coded according to heliocentric distance. The solid black lines indicate the approximate tracks of the Sagittarius leading and trailing arms. Galactic latitudes of $b=\pm 10$ deg are indicated with the dotted lines.}
    \label{fig:sdss_rrl}
\end{figure*}

We also use the QSO sample to check whether or not the proper motion uncertainties vary significantly with magnitude or colour. In Fig. \ref{fig:qso_mag_col} we show the dispersion in QSO proper motions as a function of $r$-band magnitude (left panel) and $g-r$ colour (right panel). The dotted lines indicate the median standard deviation in proper motion of 2 mas/yr. There is a weak dependence on r-band magnitude, whereby the QSO proper motion distributions get slightly broader at fainter magnitudes. However, most of the halo stars in this work have $ r < 19$ and there is little variation at these brighter magnitudes. Finally, we find no detectable dependence of $\sigma (\mu)$ on $g-r$ colour. It is worth remarking that the stability of these proper motion measurements to changes in magnitude and colour is a testament to the astrometric stability of the improved SDSS-\textit{Gaia} catalog. 
\\
\\
\noindent
In Section \ref{sec:like} we introduce a rotating velocity ellipsoid model for the Milky Way halo stars. In order to test the effects of any systematic uncertainties in the SDSS-\textit{Gaia} proper motions, we also apply this modeling procedure to the sample of SDSS DR12 QSOs.  We adopt a ``distance'' of 20 kpc, which is the mean distance to our halo star samples, and find the best fit rotation ($\langle V_{\rm \phi} \rangle$) value. This procedure gives a best fit value of $\langle V_{\rm \phi} \rangle \sim 10$ km s$^{-1}$. Note, that if there were no systematics present, then there would be no rotation signal. In Fig. \ref{fig:qso_mjd} we showed that the median proper motions of the QSOs was $\sim 0.1$ mas/yr. Indeed, at a distance of 20 kpc, this proper motion corresponds to a velocity of 10 km s$^{-1}$. Thus, although the astrometry systematics in our SDSS-\textit{Gaia} proper motion catalog are small, at the typical distances of our halo stars we cannot robustly measure rotation signals weaker than 10 km s$^{-1}$. We discuss this point further in Section \ref{sec:res}.

In the remainder of this work, we use Eqn. \ref{eqn:sig} to define the proper motion uncertainties of our halo star samples (see below). Thus, we assume that the proper motion errors are random, independent and normally distributed with variance depending only on the time-baseline between SDSS and \textit{Gaia} measurements. Note that since we are trying to measure the \textit{centroid} of the proper motion distribution (i.e. the net rotation), rather than deconvolve it into components or measure their width, we are not very sensitive to knowing the proper motion errors precisely.

\section{Stellar Halo Stars}
\label{sec:samples}

\subsection{RR Lyrae}
\label{sec:rrl}

RR Lyrae (RRL) stars are pulsating horizontal branch stars found abundantly in the stellar halo of our Galaxy. These variable stars have a well-defined Period-Luminosity-Metallicity relation, and their distances can typically be measured with accuracies of less than 10 percent. Furthermore, RRL have bright absolute magnitudes ($M_V \sim 0.6$), so they can be detected out to large distances in relatively shallow surveys.  These low mass, old (their ages are typically in excess of 10 Gyr) stars are ideal tracers of the Galactic halo, and, indeed, RRL have been used extensively in the literature to study the stellar halo (e.g. \citealt{vivas06, watkins09, sesar10, simion14, fiorentino15}). 

In this work, we use a sample of type AB RRL stars from the Catalina Sky Survey \citep{drake13a,drake13b,torrealba15} to infer the rotation signal of the Milky Way stellar halo. This survey has amassed a large number ($N \sim 22,700$) of RRL stars over 33,000 deg$^2$ of the sky, with distances in excess of 50 kpc. The RRL sample is matched to the SDSS-\textit{Gaia} proper motion catalog by searching the nearest neighbours within $10\arcsec$. Our resulting sample contains $N=8590$ RRL stars with measured 3D positions, photometric metallicities (derived using Eqn. 7 from \citealt{torrealba15}) and proper motions. The distribution of this sample on the sky in Equatorial coordinates is shown in Fig. \ref{fig:sdss_rrl}. When evaluating the Galactic velocity components of the RRL stars, the random proper motion errors (derived in Section \ref{sec:pms}) dominate over the distance errors (typically $\sim 7\%$ see e.g. \citealt{simion14}), so we can safely ignore the RRL distance uncertainties in our analysis. Note that we have checked using mock stellar haloes from the Auriga simulation suite (see Section \ref{sec:sims}) that statistical distance uncertainties of $\sim 10\%$ make little difference to our results.

\subsection{Blue Horizontal Branch}
Blue Horizontal Branch (BHB) stars, like RRL, are an old, metal poor population used widely in the literature to study the distant halo (e.g. \citealt{xue08, deason12b}). BHBs have relatively bright absolute magnitudes ($M_g \sim 0.5$), which can be simply parametrised as a function of colour and metallicity (e.g. \citealt{deason11c, fermani13a}). However, unlike their RRL cousins, photometric samples of BHB stars are often significantly contaminated by blue straggler stars, which have similar colours but higher surface gravity. Spectroscopic samples of BHBs can circumvent this problem by using gravity sensitive indicators to separate out the contaminants (e.g. \citealt{clewley02, sirko04, xue08, deason12b}).

In this work we use the spectroscopic SEGUE sample of BHB stars compiled by \cite{xue11}. This sample was selected to be relatively ``clean'' of higher surface gravity contaminants, and has already been exploited in a number of works to study the stellar halo (e.g. \citealt{xue11, deason12a, kafle13, hattori13}). By cross-matching this sample with the SDSS-\textit{Gaia} catalog, we identify $N=4553$ BHB stars. We estimate distances to these stars using the $g-r$ colour and metallicity dependent relation derived by \cite{fermani13a}. Similarly to the RRL stars, we do not take into account the relatively small ($\sim 10\%$) distance uncertainties of the BHBs in our analysis. Our resulting BHB sample has 3D positions, 3D velocities and spectroscopic metallicity estimates.

\subsection{K Giants}
Giant stars are often a useful probe of the stellar halo, owing to their bright absolute magnitudes ($M_r \sim 1$ to $-3$), and large numbers in wide-field spectroscopic surveys (e.g. \citealt{morrison00, xue14}). Moreover, giants are one of the most common tracers of \textit{external} galaxy haloes (e.g. \citealt{gilbert06,monachesi16}).  In contrast to BHB and RRL stars, giant stars populate all metallicities in old populations. Thus,  they represent a less biased tracer of the stellar halo.

The drawback of using giant stars to trace the halo is that spectroscopic samples are required to limit contamination from dwarf stars, and the absolute magnitudes of giants are strongly dependent on colour and metallicity. Here, we use the spectroscopic sample of K giants compiled by \cite{xue14}, who derive distance moduli for each star using a probabilistic framework based on colour and metallicity. A distance modulus PDF is constructed for each star, and we use the mode of the distribution  $DM_{\rm peak}$ and interval between the 84\% and 16\% percentiles, $\Delta DM = \left(DM_{84}-DM_{16}\right)/2$, as the 1$\sigma$ uncertainty. We find $N = 5814$ K giants cross-matched with the SDSS-\textit{Gaia} proper motion sample. Thus, our resulting K giant sample has 3D positions (with distance moduli described using a Gaussian PDF), 3D velocities and spectroscopic metallicities.

\section{Sagittarius Stream}
\label{sec:sgr}
\begin{figure*}
    \centering
    \includegraphics[width=16cm, height=13.71cm]{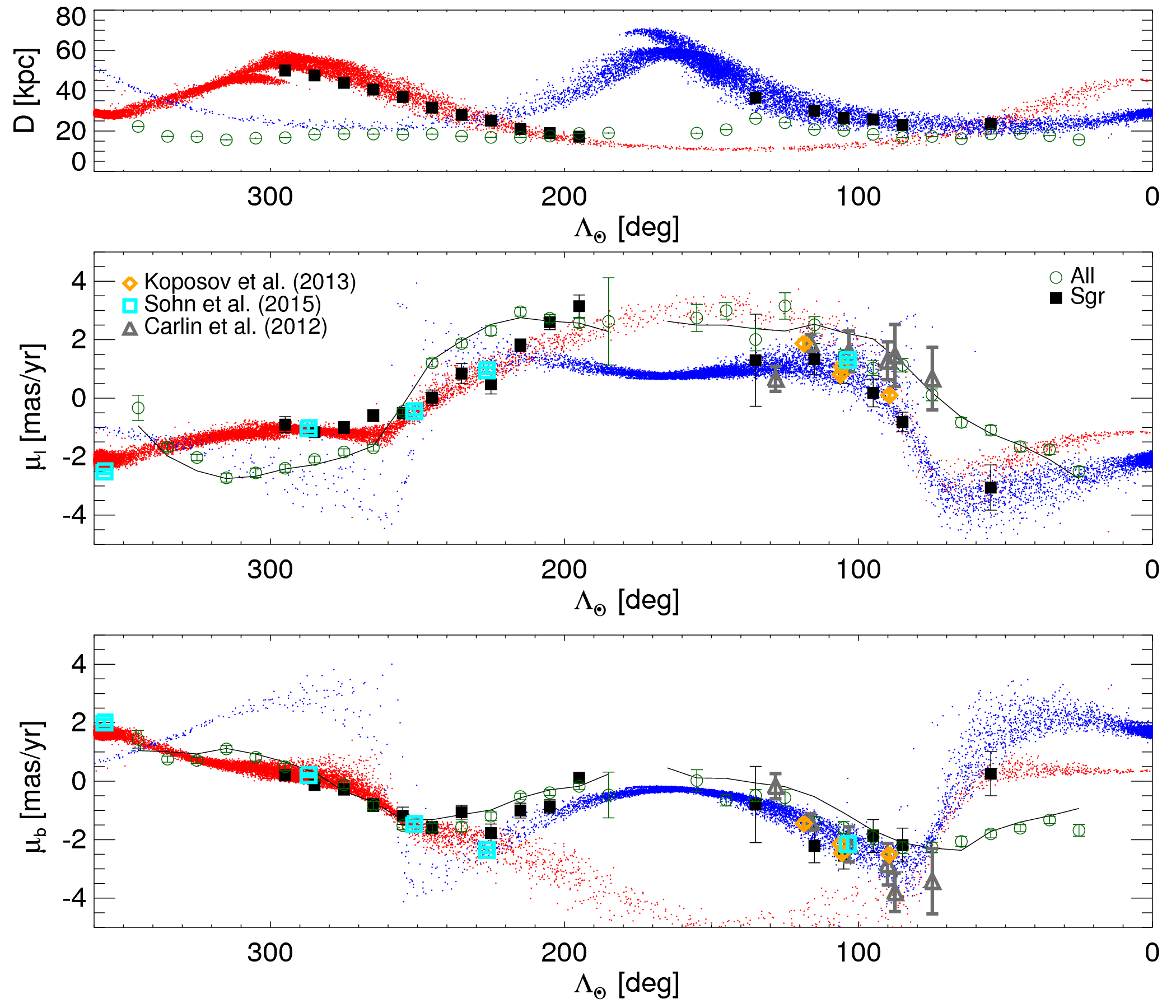}
    \caption[]{ \small Proper motions in Galactic coordinates ($\mu_l$ middle panel, $\mu_b$ bottom panel) against longitude in the Sagittarius (Sgr) coordinate system (see \citealt{majewski03}).  We also show heliocentric distance against Sgr longitude in the top panel. The red and blue points show the leading and trailing arms of the
\cite{law10} model. Note that we only show material stripped within the last 3 pericentric passages of the model orbit. The black filled squares show the median proper motions for RRL stars associated with the Sgr stream, and the open green circles are the whole sample of RRL stars.  Here, we show median proper motions in bins of $\Lambda_\odot$. The error bars indicate $1.48 \, \mathrm{MAD}/\sqrt{N}$, where MAD $=$ median absolute deviation and $N$ is the number of stars in each bin. The orange diamonds, cyan squares and grey triangles show proper motion measurements along the stream from \cite{koposov13}, \cite{sohn15,sohn16} and \cite{carlin12}.  There is excellent agreement between the SDSS-\textit{Gaia} proper motion measurements and the model values from \cite{law10} as well as previous measurements in the literature (see Fig. \ref{fig:pm_comp}). The solid black line shows the maximum likelihood model for halo rotation computed in Section \ref{sec:like}. A model with mild prograde rotation agrees very well with the proper motion data. }
    \label{fig:sgr_pms}
\end{figure*}

\begin{figure}
    \centering
    \includegraphics[width=8.5cm, height=8.5cm]{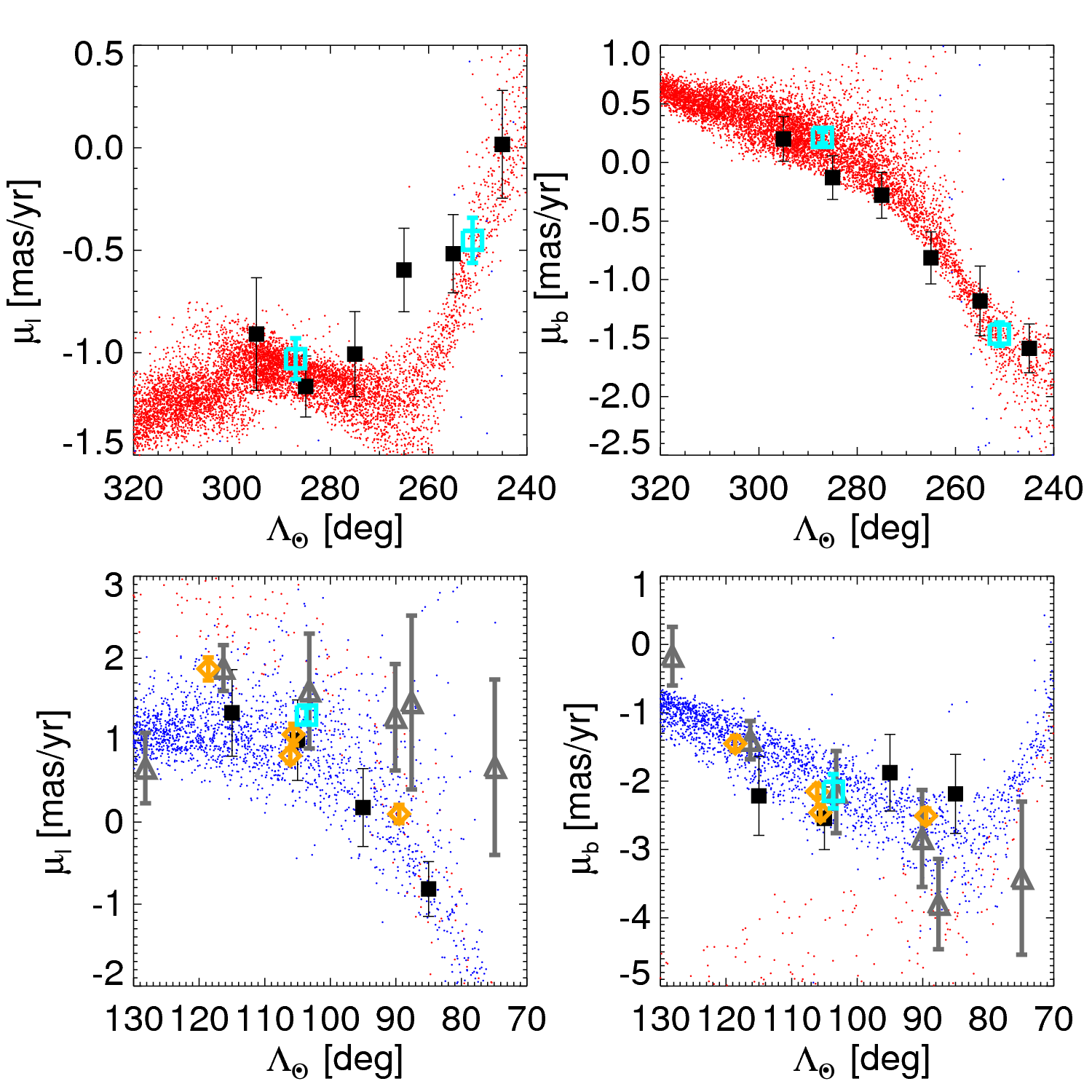}
    \caption[]{\small Proper motions in Galactic coordinates ($\mu_l$ left panels, $\mu_b$ right panel) against longitude in the Sagittarius (Sgr) coordinate system. The symbols and colors are identical to Fig. \ref{fig:sgr_pms}. Here, we have zoomed in on the regions of the Sgr stream that have proper motion constraints in the literature.  The SDSS-\textit{Gaia} RRL proper motions in the Sgr stream (solid black squares) are in excellent agreement with the literature values.}
    \label{fig:pm_comp}
\end{figure}

Before introducing our model for halo rotation, we identify RRL stars in our sample that likely belong to the Sagittarius (Sgr) stream. This vast substructure is very prominent in the SDSS footprint \citep{belokurov06}, and thus it may overwhelm any halo rotation signatures associated with earlier accretion events. Furthermore, previous works have independently measured proper motions of Sgr stars\citep{carlin12, koposov13, sohn15}, and hence we can provide a further test of our SDSS-\textit{Gaia} proper motions. Note that we use RRL stars (rather than BHBs or K giants) in Sgr as these stars have the most accurate distance measurements, and thus Sgr members can be identified relatively cleanly. 
We identify Sgr stars according to position on the sky $(\alpha,\delta)$ and heliocentric distance using the approximate stream coordinates used by \cite{deason12b} and \cite{belokurov14}. The top panel of Fig. \ref{fig:sgr_pms} shows that our distance selection of Sgr stars agrees well with the \cite{law10} model. Our selection procedure identifies $N=830$ candidate Sgr associations, which corresponds to  roughly 10\% of our RRL sample. 

In Fig. \ref{fig:sgr_pms} we show proper motions in Galactic coordinates ($\mu_\ell, \mu_b$) as a function of longitude in the Sgr coordinate system (see \citealt{majewski03}). The red and blue points show the leading and trailing arms of the \cite{law10} model of the Sgr stream. Note that we only show material stripped within the last 3 pericentric passages of the model orbit. The black filled squares show the median SDSS-\textit{Gaia} proper motions for RRL stars associated with the Sgr stream in bins of Sgr longitude, and the error bars indicate $1.48 \, \mathrm{MAD}/\sqrt{N}$, where MAD $=$ median absolute deviation and $N$ is the number of stars in each bin. It is encouraging that the Sgr stars in our RRL sample agree very well with the model predictions by \cite{law10}. Proper motion measurements of Sgr stars in the literature are also shown in Fig. \ref{fig:sgr_pms}: these are given by the orange diamonds \citep{koposov13}, cyan squares \citep{sohn15,sohn16} and grey triangles \citep{carlin12}. Our SDSS-\textit{Gaia} proper motions are in excellent agreement with these other (independent) measures (see also Fig. \ref{fig:pm_comp}). Finally, we show the proper motions for the entire sample of SDSS-\textit{Gaia} RRL stars with the open green circles. The stars associated with Sgr are clearly distinct from the overall halo in proper motion space. The solid black line shows the maximum likelihood model for halo rotation computed in Section \ref{sec:like}. A model with mild prograde rotation agrees very well with the proper motion data. Note that the variation in proper motion with $\Lambda_\odot$ in the model is largely due to the solar reflex motion. Indeed, the solar reflex motion (in proper motion space) for Sgr stars is lower because they are typically further away than the halo stars. This is the main reason for the stark difference between the proper motions of the two populations in Fig. \ref{fig:sgr_pms}.

We also show the heliocentric distances of the Sgr stars as a function of Sgr Longitude in the top panel of Fig. \ref{fig:sgr_pms}. Again, there is excellent agreement with the \cite{law10} models.  This figure shows that we can probe the Sgr proper motions out to $D \sim 50$ kpc, and thus we can accurately trace halo proper motions out to these distances (see Section \ref{sec:res}).

In Fig \ref{fig:pm_comp} we zoom in on the regions along the Sgr stream where proper motions have been measured previously in the literature. Here, the agreement with the other observational data is even clearer. In particular, our Sgr leading arm proper motions at $ 240^\circ \lesssim \Lambda_\odot \lesssim 360^\circ$ are in excellent agreement with the \textit{HST} proper motions measured by \cite{sohn15}. This is a wonderful validation of two completely independent astrometric techniques! Note that the Sgr stream is not the focus of this study, but the proper motion catalog we present here is a useful probe of the stream dynamics. For example, the slight differences between the \cite{law10} model predictions and our measurements could be used to refine/improve models of the Sgr orbit. We leave this task, and other applications of the Sgr proper motions, to a future study.
\\
\noindent
We have now shown, using both spectroscopically confirmed QSOs and stars belonging to the Sgr stream, that our SDSS-\textit{Gaia} proper motions are free of any significant systematic uncertainties. In the following Section we use this exquisite sample to infer the rotation signal of the stellar halo.

\section{Halo Rotation}
\label{sec:like}

In this Section, we use the SDSS-\textit{Gaia} sample of RRL, BHB and K giant stars to measure the average rotation of the Galactic stellar halo. Below we describe our rotating halo model, and outline our likelihood analysis. 

In order to convert observed heliocentric velocities into Galactocentric ones, we adopt a distance to the Galactic centre of $R_0=8.3 \pm 0.3$ kpc \citep{schonrich12, reid14}, and we marginalise over the uncertainty in this parameter in our analysis. Given $R_0$, the total solar azimuthal velocity in the Galactic rest frame is strongly constrained by the observed proper motion of Sgr A$^*$, i.e. $V_{g, \odot} = \mu (\mathrm{Sgr \, A^*}) \times R_0$. We adopt the \cite{reid04} proper motion measurement of Sgr A$^*$, which gives a solar azimuthal velocity of $V_{g, \odot} = 250 \pm 9$ km s$^{-1}$. Finally, we use the solar peculiar motions $(U_\odot, V_\odot, W_\odot)=(11.1, 12.24, 7.25)$ km s$^{-1}$ derived by \cite{schonrich10}. Thus, in our analysis, the circular speed at the position of the Sun is $V_c = 238$ km s$^{-1}$ (where $V_{g, \odot}= V_c +V_\odot$). We note that the combination of $R_0=8.5$ kpc and $V_c = 220$ km s$^{-1}$ has been used widely in the literature, so in Section \ref{sec:res} we show how our halo rotation signal is affected if we instead adopt these parameters.

\subsection{Model} 
We define a (rotating) 3D velocity ellipsoid aligned in spherical coordinates:

\begin{eqnarray}
\label{eqn:fv}
&&P(v_r,v_\theta,v_\phi|\sigma_r,\sigma_\phi,\sigma_\theta,\langle V_\phi \rangle) = \\ 
&&\frac{1}{\left(2\pi\right)^{3/2}\sigma_r \sigma_\theta
  \sigma_\phi} \mathrm{exp}\left[-\frac{v^2_r}{2\sigma^2_r}-\frac{v^2_\theta}{2
    \sigma^2_\theta}-\frac{\left(v_\phi-\langle V_{\phi} \rangle\right)^2}{2
    \sigma^2_\phi}\right] \notag
\end{eqnarray}
Here, we only allow net streaming motion in the $v_\phi$ velocity coordinate, and assume Gaussian velocity distributions. Note that positive $\langle V_{\phi} \rangle$ is in the same direction as the disc rotation. For simplicity, we assume an isotropic ellipsoid where $\sigma_r=\sigma_\theta=\sigma_\phi=\sigma_*$, but we have ensured that this assumption of isotropy does not significantly affect our rotation estimates (see also Section \ref{sec:sims}).

This velocity distribution function can be transformed to Galactic coordinates $(\mu_l, \mu_b, v_{\rm los})$ by using the Jacobian of the transformation $J=4.74047^2 D^2$, which gives $P(\mu_l, \mu_b,v_{\rm los}|\sigma_*,\langle V_\phi \rangle, D)$.

The RRL stars only have proper motion measurements, so, in this case, we marginalise the velocity distribution function along the line-of-sight to obtain $P(\mu_l, \mu_b|\sigma_*,\langle V_\phi \rangle)$. Furthermore, while we can safely ignore the distance uncertainties for the RRL and BHB stars, we do need to take the K giant absolute magnitude uncertainties into account (typically, $\Delta DM \sim 0.35$) . Thus, for the K giants we include a distance modulus PDF in the analysis. Here, we follow the prescription by \cite{xue14} and assume a Gaussian distance modulus distribution with mean, $\langle DM \rangle = DM_{\rm peak}$ and standard deviation, $\sigma_{DM} = \left(DM_{84}-DM_{16}\right)/2$. Here, $DM_{\rm peak}$ is the most probable distance modulus derived by \cite{xue14}, and $\left(DM_{84}-DM_{16}\right)/2$ is the central 68\% interval. This distance modulus PDF was derived by \cite{xue14} using empirically calibrated colour-luminosity fiducials, at the observed colour and metallicity of the K giants. 

\begin{eqnarray}
&&P(\mu_l, \mu_b, v_{\rm los}|\sigma_*,\langle V_\phi \rangle) = \\ 
&&\int P(\mu_l, \mu_b, v_{\rm los}|\sigma_*,\langle V_\phi \rangle, DM)
\mathcal{N}(DM|DM_0, \sigma_{DM}) d DM \notag
\end{eqnarray}
where $\mathcal{N}(DM|DM_0, \sigma_{DM})$ is the normal distribution describing the uncertainty in measuring the distance modulus to a given star.

We then use a likelihood analysis to find the best-fit $\langle V_\phi \rangle$ value. The (isotropic) dispersion, $\sigma_*$, is also a free parameter in our analysis. As we are mainly concerned with net rotation, we assume a flat prior on $\sigma_*$ in the range $\sigma_* =[50,200]$ km s$^{-1}$, and marginalise over this parameter to find the posterior distribution for $\langle V_\phi \rangle$.

When evaluating the likelihoods of individual stars under our model we also take into account the Gaussian uncertainties on proper motions as prescribed by Eq.~ \ref{eqn:sig}. As the likelihood functions are normal distributions, this amounts to a simple convolution operation.

\subsection{Results}
\label{sec:res}

\begin{table*}
\begin{center}
\renewcommand{\tabcolsep}{0.8cm}
\renewcommand{\arraystretch}{1.3}
  \begin{tabular}{c c c c c}
    \hline
    \hline
    &\multicolumn{4}{c}{\textbf{RRL}}\\
    & \multicolumn{2}{c}{All} & \multicolumn{2}{c}{Exc. Sgr} \\
    & N & $\langle V_\phi \rangle$ [km s$^{-1}$] & N & $\langle V_\phi \rangle$ [km s$^{-1}$]\\
    \cline{2-5} \\
     All $\mathrm{[Fe/H]}$ & 7456 & $12^{+2}_{-3} $ & 6663 & $9^{+3}_{-2}$ \\
    $\mathrm{[Fe/H]} >-1.5$ & 4322 & $14^{+3}_{-4} $ & 3983 & $11^{+4}_{-3}$ \\
    $\mathrm{[Fe/H]} <-1.5$ & 1460 & $10^{+6}_{-7} $ & 1312 & $6.0^{+7}_{-6}$ \\
    \hline
    \hline
    &\multicolumn{4}{c}{\textbf{BHB}}\\
    & \multicolumn{2}{c}{All} & \multicolumn{2}{c}{Exc. Sgr} \\
    & N & $\langle V_\phi \rangle$ [km s$^{-1}$]& N & $\langle V_\phi \rangle$ [km s$^{-1}$]\\
    \cline{2-5} \\
     All $\mathrm{[Fe/H]}$& 3947 & $6.0^{+3}_{-3} $ & 3671 & $5.0^{+3}_{-3}$ \\
    $\mathrm{[Fe/H]} >-1.5$ & 756 & $18^{+7}_{-7} $ & 715 & $21^{+7}_{-7}$ \\
    $\mathrm{[Fe/H]} <-1.5$ & 3191 & $2.0^{+4}_{-3}$ & 2956 & $0.0^{+4}_{-3}$ \\
    $\mathrm{[Fe/H]} >-1.5$, PM only & 756 & $15^{+8}_{-9}$ & 715 & $19^{+8}_{-9}$ \\
    $\mathrm{[Fe/H]} <-1.5$, PM only & 3191 & $1.0^{+4}_{-4}$ & 2956 & $-1.0^{+4}_{-4}$ \\
    \hline
    \hline
    &\multicolumn{4}{c}{\textbf{K giants}}\\
    & \multicolumn{2}{c}{All} & \multicolumn{2}{c}{Exc. Sgr} \\
    & N & $\langle V_\phi \rangle$ [km s$^{-1}$] & N & $\langle V_\phi \rangle$ [km s$^{-1}$] \\
    \cline{2-5} \\
     All $\mathrm{[Fe/H]}$& 5284 & $23^{+3}_{-3} $ & 4603 & $19^{+3}_{-3}$  \\
    $\mathrm{[Fe/H]} >-1.5$ & 2553 & $28^{+4}_{-4}$ & 2159 & $23^{+4}_{-4}$ \\
    $\mathrm{[Fe/H]} <-1.5$ & 2731 & $17^{+4}_{-4}$ & 2444 & $14^{+4}_{-4}$ \\
     $\mathrm{[Fe/H]} >-1.5$, $P_{\rm RGB} > 0.8$ & 1748 & $30^{+5}_{-5}$ &  1426 & $23^{+5}_{-5}$ \\
    $\mathrm{[Fe/H]} <-1.5$, $P_{\rm RGB} > 0.8$ & 1985 & $22^{+5}_{-5}$ & 1744 & $18^{+5}_{-5}$ \\
    \hline
    \hline
    \label{tab:res}
    \end{tabular}
    \caption{\small A summary of best-fit $\langle V_\phi \rangle$ values and associated $1 \sigma$ uncertainties. Halo stars with $r < 50$ kpc, $|z| > 4$ kpc and $\mu < 100$ mas/yr were used to derive these values.}
\end{center}
\end{table*}

In this Section, we apply our likelihood procedure to RRL, BHB and K giant stars with SDSS-\textit{Gaia} proper motions. For all halo tracers, we only consider stars with $r < 50$ kpc and $|z| > 4$ kpc. The latter cut is imposed to avoid potential disc stars. In addition, we remove any stars with considerable proper motion ($\mu > 100$ mas/yr), although, in practice this amounts to removing only a handful ($\ll 1\%$) of stars and their exclusion does not affect our rotation estimates. The best fit values of $\langle V_\phi \rangle$ described in this section are summarised in Table \ref{tab:res}.

\begin{figure}
  \centering
  \includegraphics[width=8.5cm, height=4.25cm]{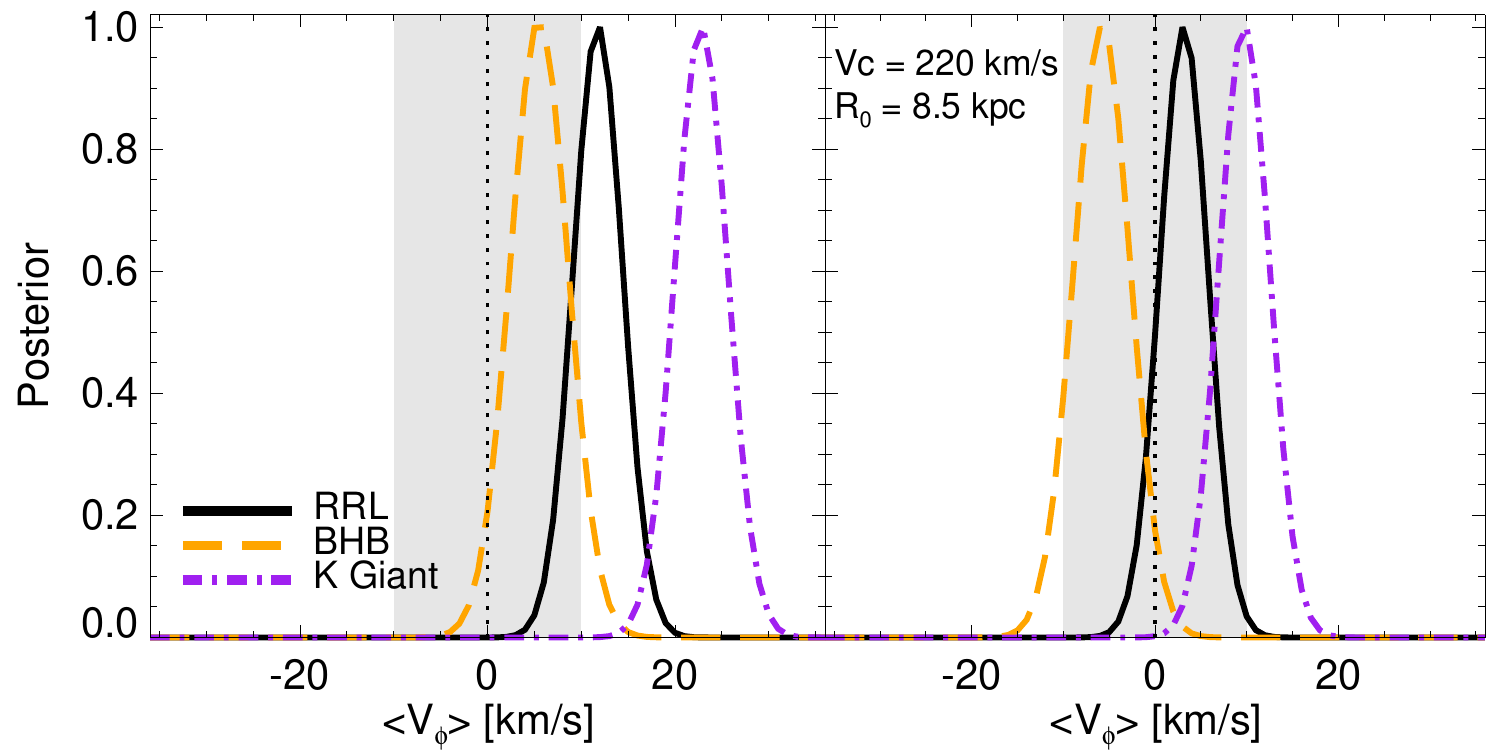}
  \caption{ \small The posterior $\langle V_\phi \rangle$ distributions for RRL (solid black), BHB (dashed orange) and K giant (dot-dashed purple) tracers. The shaded grey region indicates the approximate systematic uncertainty in the proper motion measurements ($\sim 10$ km s$^{-1}$ at $D=20$ kpc). For comparison, the right-hand panel shows the posterior distributions when a different combination of position of the Sun ($R_0=8.5$ kpc), and circular velocity at the position of the Sun ($V_c = 220$ km s$^{-1}$) is used. With a lower solar azimuthal velocity, the (already mild) rotation signal disappears. Current estimates favour the larger value of $\sim 240$ km s$^{-1}$, but it is worth bearing in mind the degeneracy between the rotation signal and adopted solar motion.}
  \label{fig:vphi}
\end{figure}

In Fig. \ref{fig:vphi} we show the posterior distribution for $\langle V_\phi \rangle$ for each of the halo tracers. The solid black, dashed orange and dot-dashed purple lines show the results for RRL, BHBs and K giants, respectively. All the halo tracers favour a mild prograde rotation signal, with $\langle V_\phi \rangle \sim 5-25$ km s$^{-1}$. Note that the RRL model is shown against the proper motion data in Fig. \ref{fig:sgr_pms}. In general, the K giants show the strongest rotation signal of the three halo tracers. This is likely because the K giants have a broader age and metallicity spread than the RRL and BHB stars (see Section \ref{sec:sims}). However, the K giant rotation signal is still relatively mild ($\sim 20$ km s$^{-1}$) and similar (within 10-15 km s$^{-1}$) to the RRL and BHB results. The three tracer populations have different distance distributions, so it is not immediately obvious that their rotation signals can be directly compared. However, as we show in Fig. \ref{fig:vphi_rad}, we find little variation in the rotation signal with Galactocentric radius, so a comparison between the ``average'' rotation signal of the populations is reasonable. Finally, we note that we also check that the Sgr stars in our sample make little difference to the overall rotation signal of the halo (see Table \ref{tab:res}). 

For comparison, the right-hand panel of Fig. \ref{fig:vphi} shows the posterior  distributions if we adopt other commonly used parameters for distance from the Galactic centre and circular velocity at the position of the Sun: $R_0=8.5$ kpc, $V_c = 220$ km s$^{-1}$. In this case, only the K giants exhibit a detectable rotation signal. It is worth emphasizing that current estimates of the solar azimuthal velocity favour the larger value of $V_c \sim 240$ km s$^{-1}$ \citep{bovy12,schonrich12,reid14} that we use, but it is important to keep in mind that the rotation signal is degenerate with the adopted solar motion. In addition, as discussed in Section \ref{sec:pms}, the systematic uncertainties of our SDSS-\textit{Gaia} proper motion catalogue are at the level of $\sim 0.1$ mas/yr. Thus, for typical distances to the halo stars of 20 kpc, we cannot robustly measure a rotation signal that is weaker than 10 km s$^{-1}$.

\begin{figure*}
  \centering
  \includegraphics[width=16cm, height=3.64cm]{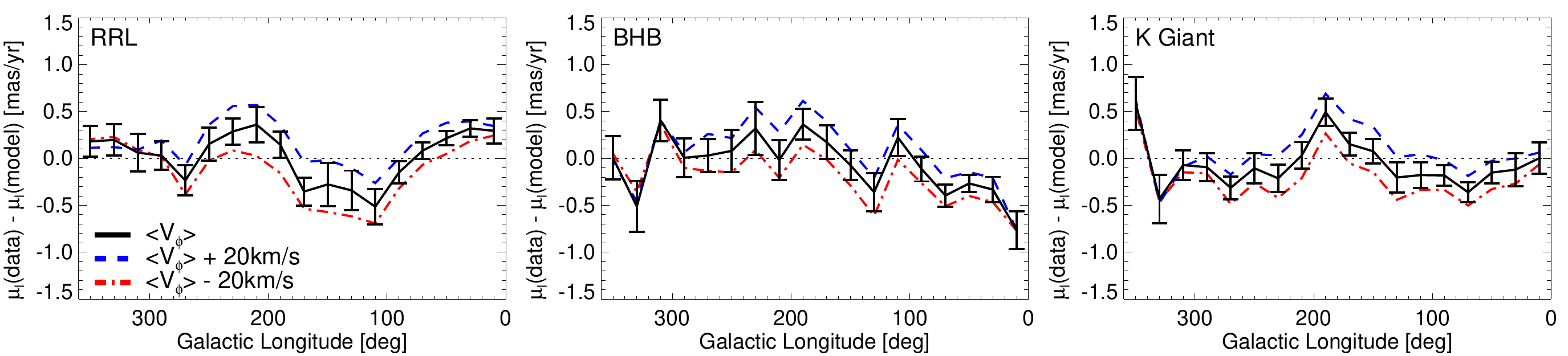}
  \caption{ \small A comparison between the observed proper motions and the maximum likelihood model predictions. Here, we show the median $\mu_l$ values in bins of Galactocentric longitude. The RRL, BHB and K giant samples are shown in the left, middle and right panels, respectively. The solid black line shows the comparison with the maximum likelihood model. The error bars show the median absolute deviation of the data in each Galactic longitude bin. The dashed blue and dot-dashed red lines show models with $\langle V_\phi \rangle \pm 20$ km s$^{-1}$. Note that in this comparison we have removed stars that likely belong to the Sgr stream.}
  \label{fig:dm}
\end{figure*}

\begin{figure*}
    \centering
    \includegraphics[width=17cm, height=4.25cm]{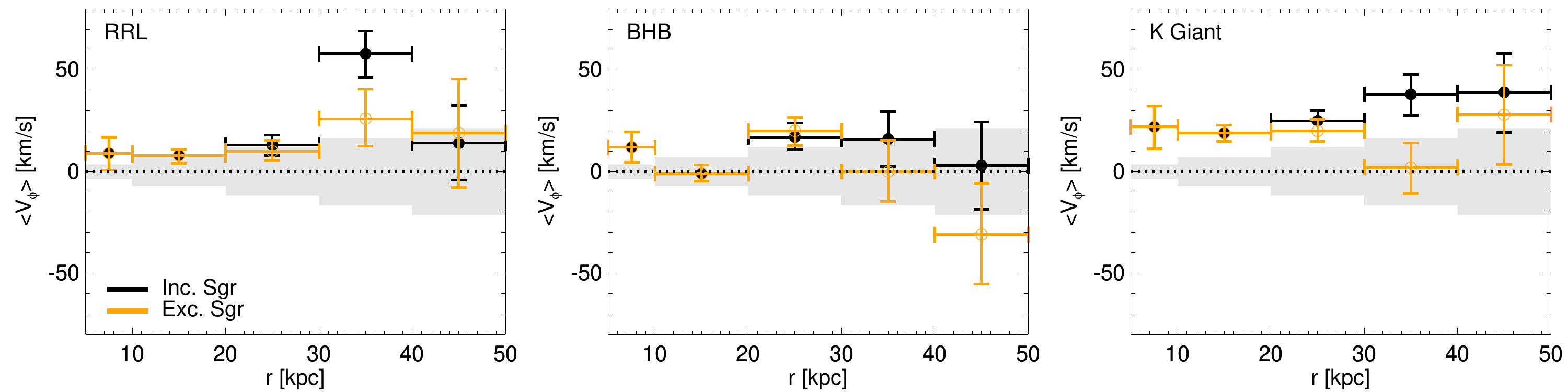}
    \caption{\small The best-fit rotation of the stellar halo, $\langle V_\phi \rangle$, in Galactocentric radial bins. Here, 10 kpc radial bins are used and the error bars indicate the 1$\sigma$ confidence levels. RRL, BHBs and K giants are shown in the left, middle and right panels, respectively. The solid black circles show all halo stars, and the open orange circles show the rotation signal when stars likely associated with the Sagittarius (Sgr) stream are removed.  The apparent rotation signal at radii $ 30 < r/\mathrm{kpc} < 40$ is due to the Sgr stream. The shaded grey region indicates the approximate systematic uncertainty of the SDSS-\textit{Gaia} proper motions ($\sim 0.1$ mas/yr).}
    \label{fig:vphi_rad}
\end{figure*}

In Fig. \ref{fig:dm} we compare the model predictions for $\mu_l$ with the observed data. We show the Galactic longitude proper motion $\mu_l$ because this component is more sensitive than $\mu_b$ to variations in $\langle V_\phi \rangle$. The solid black line shows the difference between the maximum likelihood models and the data as a function of Galactocentric longitude. The error bars indicate the median absolute deviation of the data in each bin. For comparison, we also show with the dashed blue and dot-dashed red lines the model predictions with $\langle V_\phi \rangle \pm 20$ km s$^{-1}$. For all three tracers, the models with very mild prograde rotation agree well with the data.

Our maximum likelihood models give $\sigma^*$ values of 138, 121 and 111 km s$^{-1}$ for the RRL, BHBs and K giants respectively. These values agree well with previous estimates of in the literature \citep{battaglia05, brown10, deason12b, xue08}. Note that our models assume isotropy, but we find that both radially and tangentially biased models make little difference to our estimates of $\langle V_\phi \rangle$.

\begin{figure}
  \centering
  \includegraphics[width=8cm, height=12cm]{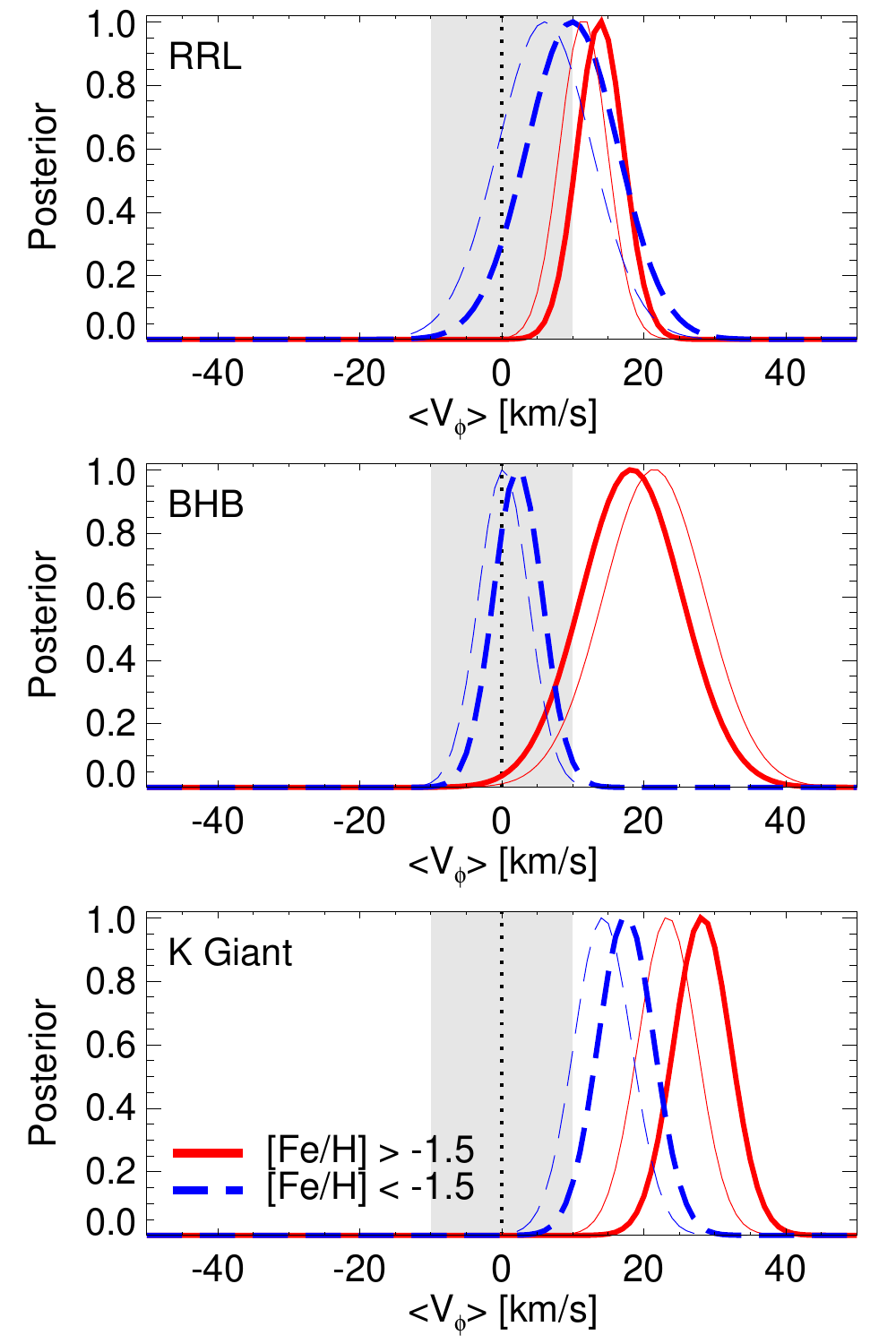}
  \caption{ \small The posterior $\langle V_\phi \rangle$ distributions for metal-richer (solid red lines, [Fe/H] $>-1.5$) and metal poorer (dashed blue lines, [Fe/H] $<-1.5$). RRL, BHBs and K giants are shown in the top, middle and bottom panels, respectively. The metal-richer BHB and K giant stars are mildly biased ($\sim 1\sigma$) towards stronger prograde rotation. The thinner lines show the estimated rotation signals when stars associated with the Sgr stream are excluded. The shaded grey region indicates the approximate systematic uncertainty in the proper motion measurements ($\sim 10$ km s$^{-1}$ at $D=20$ kpc) }
  \label{fig:vphi_met}
\end{figure}

We now investigate if there is a radial dependence on the rotation signal of the stellar halo. Our likelihood analysis is applied to halo stars in radial bins 10 kpc wide between Galactocentric radii $0 < r/\mathrm{kpc} < 50$. The results of this exercise are shown in Fig. \ref{fig:vphi_rad}. The solid black circles show all halo stars, and the open orange circles show the rotation signal when stars likely associated with the Sagittarius (Sgr) stream are removed. Here, the error bars indicate the 1$\sigma$ confidence levels. We find that the (prograde) rotation signal stays roughly constant at $10 \lesssim \langle V_\phi/ \mathrm{km \, s^{-1}} \rangle \lesssim 20$. We do find a stronger rotation signal in the radial bin $ 30 < r/ \mathrm{kpc} < 40$ for both RRL and K giants, but this is attributed to a significant number of Sgr stars in this radial regime. The shaded grey regions in Fig. \ref{fig:vphi_rad} indicate the approximate \textit{systematic} uncertainty in the velocity measurements in each radial bin, assuming a systematic proper motion uncertainty of 0.1 mas/yr. Thus, the prograde rotation is very mild, and we are only just able to discern a rotation signal that is not consistent with zero. 

In Fig. \ref{fig:vphi_met} we explore whether or not the rotation signal of the halo stars is correlated with metallicity. The spectroscopic BHB and K giant samples have measured [Fe/H] values, and for the RRL we use photometric metallicities measured from the light curves. The metallicity distribution functions of the three halo tracers are different, and we are using both spectroscopic and photometric metallicities. Thus, we only compare ``metal-richer'' and ``metal-poorer'' stars using a metallicity boundary of [Fe/H] $=-1.5$. This boundary was chosen as the median value of the K giant sample, which is the least (metallicity) biased tracer. In Fig. \ref{fig:vphi_met} we show the posterior probability distributions for the average rotation of the metal-rich (solid red) and metal-poor (dashed blue) tracers. The thinner lines show the posteriors when stars likely associated with the Sgr stream are excluded. There is no evidence for a metallicity dependence in the RRL sample, but both the BHBs and K giants show a slight ($\sim 1\sigma$) bias towards stronger prograde rotation for metal-rich stars. 

The lack of a metallicity correlation in the rotation of the RRL stars could be due to the relatively poor photometric metallicity estimates (see e.g. Fig. 10 in \citealt{watkins09}), which could wash out any apparent signal. On the other hand, the apparent metallicity correlation in the BHB and K giant samples could be caused by contamination. We explore this scenario in more detail below.

Previous work using only line-of-sight velocities have also found evidence for a metal-rich/metal-poor kinematic dichotomy in spectroscopic samples of BHB stars \citep{deason11a, kafle13, hattori13}. However, \cite{fermani13b} argue that this signal is due to (1) contamination by blue straggler stars, (2) incorrect distance estimates and, (3) potential pipeline systematics in the \cite{xue11} BHB sample. The BHB sample used in this work should not suffer from significant blue straggler (or main sequence star) contamination. Moreover, our distance calibration is robust to systematic metallicity differences \citep{fermani13a}. However, we cannot ignore the potential line-of-sight velocity systematics in the \cite{xue11} sample. \cite{fermani13b} find that a subsample of hot metal-poor BHB stars exhibit peculiar line-of-sight kinematics, which likely causes the metallicity bias in the rotation estimates. It is worth noting that the peculiar line-of-sight kinematics of the hot BHB stars could also be due to a stream-like structure in the halo, and is not necessarily a pipeline failure. In Table \ref{tab:res} we also give the rotation estimates for metal-rich/metal-poor stars computed with proper motions only. The results are only slightly changed when we do not use the BHB line-of-sight velocities, and they agree within $1 \sigma$ of the rotation estimates when 3D velocities are used. 

We also investigate whether or not the apparent metallicity correlation in the K giant sample could be due to contamination. For example, if there are (metal-rich) disc stars present in the sample this could lead to a stronger prograde signal in the metal-richer stars.  Disc contamination could result from stars being misclassified as giant branch stars (e.g. dwarfs, red clump stars) and thus their distances are overestimated. To this end, we use a stricter cut on the $P_{\rm RGB}$ parameter provided by \cite{xue14}, which gives the probability of being a red giant branch stars. Our fiducial sample has $P_{\rm RGB} > 0.5$. We find that using $P_{\rm RGB} > 0.8$ results in little difference to the rotation signal of the metal-rich stars, and the rotation signal of the metal-poor stars becomes slightly stronger (see Table \ref{tab:res}). It does not appear that the sample is contaminated by disc stars, but the (slight) metallicity correlation in the K giant sample does lose statistical significance if a stricter cut on red giant branch classification is used. However, this is likely because the error bars are inflated due to smaller number statistics.

It is worth noting that the tests we perform above on the BHB and K giant samples do not significantly change the rotation signals of the stars (differences are less than $1 \sigma$), so, we are confident that contamination in these samples is not significantly affecting our results. Thus, we conclude that there does appear to be a mild correlation between rotation signal and metallicity in the halo star kinematics. 

In summary, we find that the (old) stellar halo, as traced by RRL, BHB and K giant stars, has a very mild prograde rotation signal, and there is a weak correlation between rotation signal and metallicity.  Is this the expected result for a Milky Way-mass galaxy stellar halo? Or, indeed, is this rotation signal result consistent with the predictions of the $\Lambda\mathrm{CDM}$ model? In the following Section, we exploit a suite of state-of-the-art cosmological simulations in order to address these questions.

\section{Simulated Stellar Haloes}
\label{sec:sims}

\subsection{Auriga Simulations}
\begin{figure*}
    \centering
    \includegraphics[width=17cm, height=5.1cm]{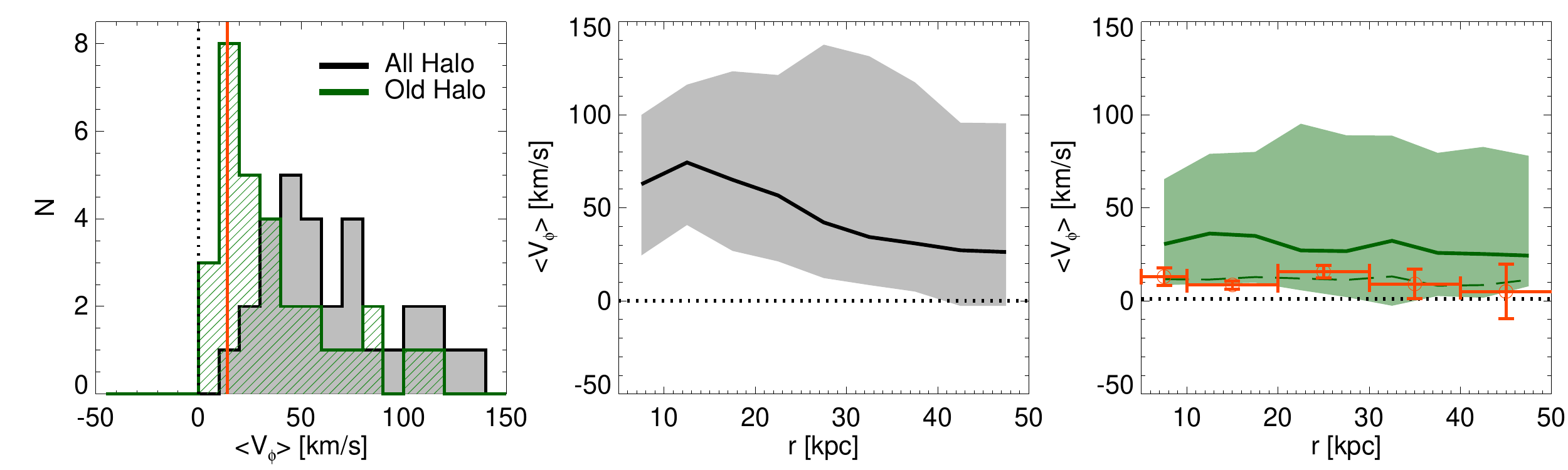}
    \caption{ \small \textit{Left panel:} The distribution of average azimuthal velocity ($\langle V_\phi \rangle $) of the 30 Auriga stellar haloes. Halo stars are spatially selected in the SDSS survey footprint with Galactocentric radius $5 < r/\mathrm{kpc} < 50$ and height above disc plane $|z| > 4$ kpc. The solid grey histogram shows all halo stars, and the line-filled green histogram shows old halo stars selected with $T_{\rm form} > 10$ Gyr. These ``old'' star particles are selected for a more direct comparison with the observed halo stars. The haloes have a range of rotation amplitudes between $0 \lesssim \langle V_{\phi} \rangle /\mathrm{km \, s^{-1}} \lesssim 120$. Old halo stars typically show a milder rotation signal with $0 \lesssim \langle V_{\phi} \rangle/\mathrm{km \, s^{-1}} \lesssim 80$. The solid red line indicates the approximate average rotation signal ($\sim 14$ km s$^{-1}$) of the old Milky Way halo populations. \textit{Middle and right panels:} The variation of rotation signal with Galactocentric radius. The solid black line and the grey filled region (middle panel) shows the median and 10th/90th percentile range for the 30 Auriga haloes. The solid green line and green filed region (right panel) are the same for the old halo stars. For comparison, we show the average (inverse variance weighted) rotation signal for RRL, BHB and K giant stars in the Milky Way with the red symbols (cf. Fig. \ref{fig:vphi_rad}). The mild prograde rotation we find in the observational data is in good agreement with the old halo stars in the simulations. The dashed green line indicates the 20th percentile of the distribution, which approximately follows the observed radial trend.}
    \label{fig:sim_rot}
\end{figure*}

In this Section, we use a sample of $N=30$ high-resolution Milky Way-mass haloes from the Auriga simulation suite. These simulations are described in more detail in \cite{grand17}, and we only provide a brief description here.

A low-resolution dark matter only simulation with box size 100 Mpc $h^{-1}$ was used to select candidate Milky Way-mass ($1 < M_{200}/10^{12}M_\odot < 2$) haloes. These candidate haloes were chosen to be relatively isolated at $z=0$. More precisely, there are no objects with masses greater than half of the parent halo closer than 1.37 Mpc. A $\Lambda$CDM cosmology consistent with the \cite{planck14} data release is adopted with parameters, $\Omega_m=0.307$, $\Omega_b=0.048$, $\Omega_\Lambda=0.693$ and $H_0=100 \, h$ km s$^{-1} $Mpc$^{-1}$, where $h=0.6777$. Each candidate halo was re-simulated at a higher resolution using a multi-mass particle ``zoom-in'' technique.

The zoom re-simulations were performed with the state-of-the-art cosmological magento-hydrodynamical code \textsc{arepo} \citep{springel10}. Gas was added to the initial conditions by adopting the same technique described in \cite{marinacci14a, marinacci14b}, and its evolution was followed by solving the MHD equations on a Voronoi mesh. At the resolution level used in this work (level 4), the typical mass of a dark matter particle is $3 \times 10^5M_\odot$, and the baryonic mass resolution is $5 \times 10^4M_\odot$. The softening length of the dark matter particles and star particles grows with time in physical space until a maximum of 369 pc is reached at $z=1.0$ (where z is the redshift). The gas cells have a softening length that scales with the mean radius of the cell, and the maximum physical softening is 1.85kpc.

The Auriga simulations employ a model for galaxy formation physics that includes critical physical processes, such as star formation, gas heating/cooling, feedback from stars, metal enrichment, magnetic fields, and the growth of supermassive black holes (see \citealt{grand17} for more details). The simulations have been successful in reproducing a number of observable disc galaxy properties, such as rotation curves, star formation rates, stellar masses, sizes and metallicities.

This work is concerned with stellar haloes of the Auriga galaxies. A future study (Monachesi et al. in preparation) will present a more general analysis of the simulated stellar halo properties. Here, we focus on the net rotation of the Auriga stellar haloes for comparison with the observational results in the preceding sections.

\subsection{Rotation of Auriga Stellar Haloes}

The definition of ``halo stars'', in both observations and simulations, is somewhat arbitrary, and often varies widely between different studies. In this work, for a more direct comparison with our observational results, we spatially select stars within the SDSS survey footprint (see Fig. \ref{fig:sdss_rrl}) with Galactocentric radius $5 < r/\mathrm{kpc} < 50$ and height above disc plane $|z| > 4$ kpc. Note that the scale heights of the Auriga discs are generally thicker than the Milky Way disc (see \citealt{grand17}), so our spatial selection will likely include some disc star particles, particularly at small radii. Finally, for a fair comparison with the old halo tracers (i.e. RRL, BHBs, and K giants) used in this work, we also select ``old'' star particles. For this purpose, we consider halo stars that formed more than 10 Gyr ago in the simulations. Note that we align each halo with the stellar disc angular momentum vector, which we compute using all star particles within 20 kpc.

In the left-hand panel of Fig. \ref{fig:sim_rot} we show the distribution of average azimuthal velocity ($\langle V_\phi \rangle$) of halo stars in the 30 Auriga simulations. Here, halo stars are selected within the SDSS survey footprint between 5 and 50 kpc from the Galactic centre, and with height above the disc plane, $|z| > 4$ kpc. The average rotation for all halo stars in this radial range are shown with the grey histogram. Old halo stars (with $T_{\rm form} >$ 10 Gyr) are shown with the green line-filled histogram. The stellar haloes show a broad range of rotation velocities, ranging from $0 \lesssim \langle V_\phi \rangle /\mathrm{km \, s}^{-1} \lesssim 120$, but they are all generally \textit{prograde}. Similarly, the old halo stars exhibit prograde rotation, but they have much milder rotation amplitudes, with $\langle V_\phi \rangle \lesssim 80$ km s$^{-1}$. The average rotation signal of the three Milky Way halo populations we used in Section \ref{sec:res} is $14$ km s$^{-1}$. Only 3 percent of the Auriga haloes have net rotation signals $\le 14$ km s$^{-1}$, however, the fraction of "old" simulated haloes with similarly low rotation amplitudes is higher (20 percent).

In the middle- and right-hand panels we show the radial dependence of the rotation signal in the simulations. Here, in the middle panel, the solid black line shows the median value of the 30 Auriga haloes and the grey shaded region indicates the 10th/90th percentiles. Similarly, in the right-hand panel, the solid green line shows the median value of the old halo stars and the green shaded region indicates the 10th/90th percentiles. The rotation signal of the whole halo sample varies with radius and declines from $\langle V_\phi \rangle \sim 70$ km s$^{-1}$ at $r \sim 10$ kpc to  $\langle V_\phi \rangle \sim 25$ km s$^{-1}$ at $r \sim 50$ kpc. In contrast, the old halo stars have a fairly constant rotation amplitude with Galactocentric distance of 20-30 km s$^{-1}$. It is likely that the higher rotation amplitude for halo stars at small Galactocentric distances is due to disc contamination and/or the presence of \textit{in situ} stellar halo populations more akin to a ``thick disc'' component (e.g. \citealt{zolotov09, font11, mccarthy12, pillepich15}). However, the old halo stars will suffer much less contamination from the disc (or disc-like) populations\footnote{It is worth noting that not \textit{all} old stars will have an external origin, as there are old ($T_{\rm form} > 10$ Gyr) populations present in the disc and \textit{in situ} halo components (see e.g. \citealt{mccarthy12, pillepich15}).}, and they are dominated by stellar populations accreted from dwarf galaxies (see e.g Figure 10 in \citealt{mccarthy12}). This is likely the reason why the rotation amplitude of the old halo stars is fairly constant with Galactocentric radius. 

Finally, it is worth noting that there are a significant number of the Auriga galaxies ($\sim 1/3$) that have an ``\textit{ex situ} disc'' formed from massive accreted satellites \cite{gomez17b}. Some of these \textit{ex situ} discs can extend more than 4 kpc above the disc plane, and can be the cause of significant rotation in the stellar halos. However,  this is not true for all of the \textit{ex situ} discs in the simulations: some are largely confined to small $|z|$ and will not necessarily affect the rotation signal at the larger Galacocentric radii probed in this work (see \citealt{gomez17b}).

We also show our observational results from the RRL, BHB and K giant stars in Fig. \ref{fig:sim_rot} (cf. Fig. \ref{fig:vphi_rad}). Here, we show the average (inverse variance weighted) rotation signal from the three populations. In practice, the rotation of the three populations is very similar (see Fig. \ref{fig:vphi_rad}). The observed rotation amplitude in the Galactic halo broadly agrees with the old halo population in the simulations: a mild prograde signal is consistent, and indeed \textit{typical} in the cosmological simulations. The dashed green line in the right-hand panel of Fig. \ref{fig:sim_rot} indicates the 20th percentile level, which agrees well with the observed values. Thus, while the mild prograde rotation of the old Milky Way halo stars is consistent with the simulated haloes, the observed rotation amplitude is on the low side of the distribution of Auriga haloes. We note that the Auriga haloes are randomly selected from the most isolated quartile of haloes (in the mass range $1-2 \times 10^{12}M_\odot$), and thus they are typical field disc galaxies (as opposed to those in a cluster environment). Thus, we can infer that the rotation signal of the old Milky Way halo is fairly low compared to the general field disc galaxy population.

It is not immediately obvious why the old halo stars in the simulations, even out to $r \sim 50$ kpc, have (mild) prograde orbits. If most of these stars come from destroyed dwarf galaxies, then their net spin will be related to the original angular momentum vectors of the accreted dwarfs. Previous studies using cosmological simulations have shown that subhalo accretion is anisotropic along filamentary structures, and is generally biased along the major axis of the host dark matter halo \citep{bailin05b, libeskind05, zentner05}. Indeed, \cite{lovell11} showed that the subhalo orbits in the Aquarius simulations are mainly aligned with the main halo spin. Hydrodynamic simulations predict that the angular momentum vector of disc galaxies tends to be aligned with the dark matter halo spin, at least in the inner parts of haloes (e.g. \citealt{bailin05, bett10, deason11b, shao16}). Thus, the slight preference for prograde orbits in the accreted stellar haloes is likely due to the filamentary accretion of subhaloes, which tend to align with the host halo major axis and stellar disc. Note that the non-perfect alignment between filaments, dark matter haloes and stellar discs will naturally lead to a relatively weak (but non-zero!) signal. In addition, the orbital angular momentum of massive accreted satellites can align with the host disc angular momentum \textit{after} infall. Indeed, \cite{gomez17b} show that when \textit{ex situ} discs are formed from the accretion of massive satellites the angular momentum of the dwarfs can be initially misaligned with the disc but can rapidly become aligned after infall. Furthermore, this alignment is not just due to a change in the satellite orbit, but also because of a response of the host galactic disc!

Note that, as mentioned above, some of the old stars will also belong to the \textit{in situ} halo component, which are more likely biased towards prograde (or disc-like) orbits. Thus, it is likely that those haloes with minor net rotation are less dominated by in situ populations. Indeed, the mild prograde rotation we see in the observational samples suggests that the in situ component of the Milky Way is relatively minor. Moreover, as more recent, massive mergers will lead to higher net spin in the halo, the weak rotation signal in the Milky Way halo is indicative of a quiescent merger history (see e.g. \citealt{gardner01, vitvitska02}).

\begin{figure}
  \centering
  \includegraphics[width=8.5cm, height=7.08cm]{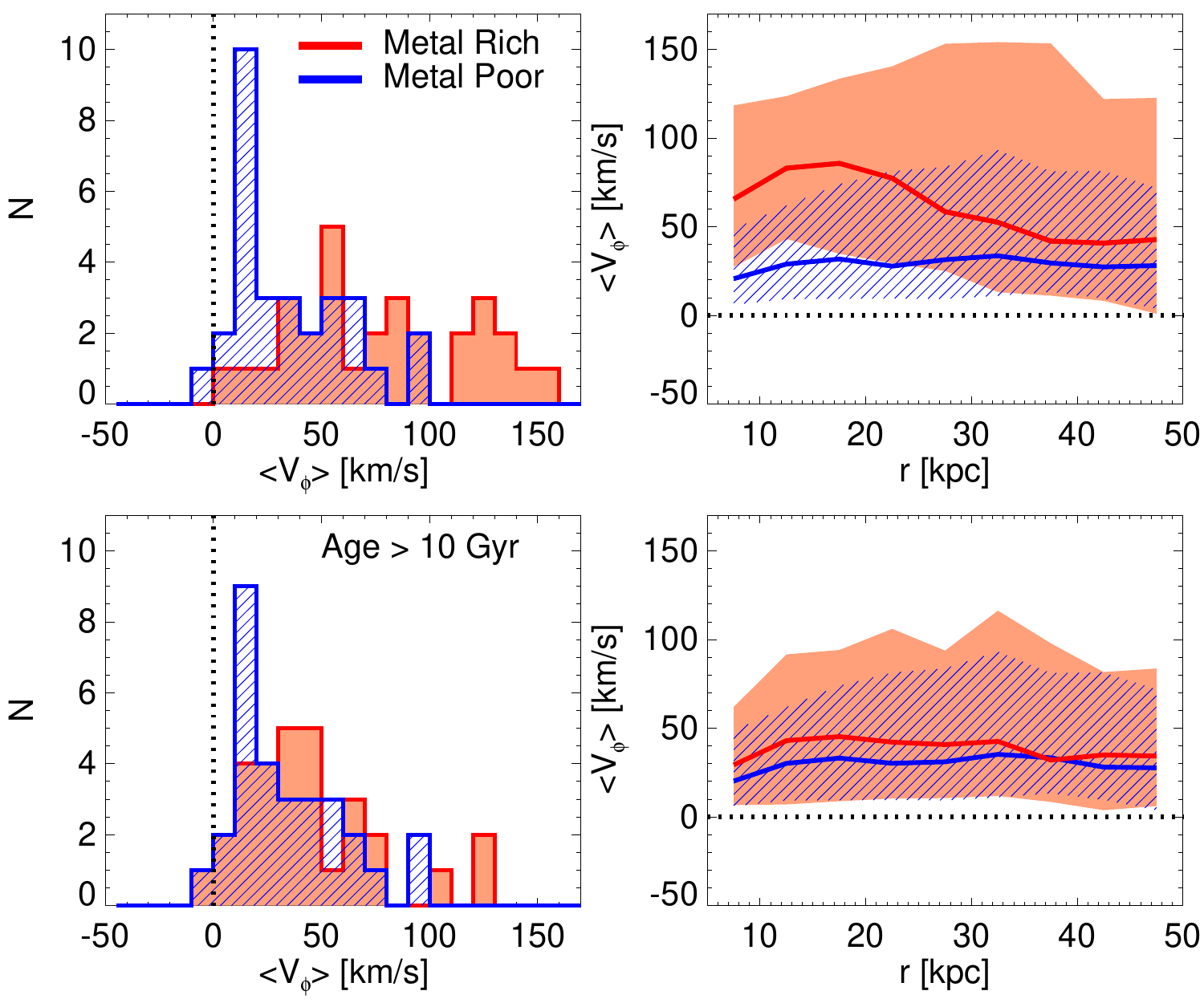}
  \caption{ \small \textit{Left panels: } The distribution of $\langle V_\phi \rangle$ of halo stars in the Auriga simulations that are metal-rich ([Fe/H] $>-1$, red) and metal-poor ([Fe/H] $<-1$, blue). All halo stars with Galactocentric radius $5 < r/\mathrm{kpc} < 50$ and height above disc plane $|z| > 4$ kpc, are shown in the top panels, and old halo stars ($T_{\rm form} > 10$ Gyr) are shown in the bottom panels. \textit{Right panels:} The variation of $\langle V_\phi \rangle$ with Galactocentric radius. Solid red (metal-rich) and line-filled blue (metal-poor) regions indicate the 10th/90th percentile ranges for the 30 Auriga haloes. The median values are shown with the solid lines.  Metal-richer stars tend to have a stronger (prograde) rotation signal than the metal-poorer stars. However, the old halo stars show a much milder metallicity correlation.}
  \label{fig:sim_rot_met}
\end{figure}

In Fig. \ref{fig:sim_rot_met} we show how the rotation signal of the Auriga stellar haloes depends on metallicity. We define ``metal-rich and ``metal-poor'' populations as halo stars with metallicities above/below 1/10th of solar ([Fe/H] $= -1$). This metallicity boundary was chosen as it roughly corresponds to the median metallicity of the old halo stars in the simulations. However, as is the case in the observations, our choice of metallicity boundary is fairly arbitrary. When all halo stars are considered, there is a tendency for the metal-richer stars to have a stronger prograde rotation. This metallicity correlation is more prominent in the inner regions of the halo. It's likely that the correlation in the inner regions of the halo is, at least in part, attributed to disc contamination and/or the presence of \textit{in situ} (disc-like) stellar halo populations. Furthermore, most of the strongly rotating \textit{ex situ} disc material in the simulations is contributed by one massive, and thus metal-rich, satellite, which could also cause a metallicity correlation in the halo stars. The old halo stars, which suffer less from disc contamination,  show only a very mild ($\sim 5-10$ km s$^{-1}$) bias towards more strongly rotating metal-rich populations. Indeed, we found a weak metallicity correlation in the observed samples of old halo stars, which seems to be in good agreement with the predictions of the simulations.

\begin{figure*}
    \centering
    \includegraphics[width=17cm, height=4.25cm]{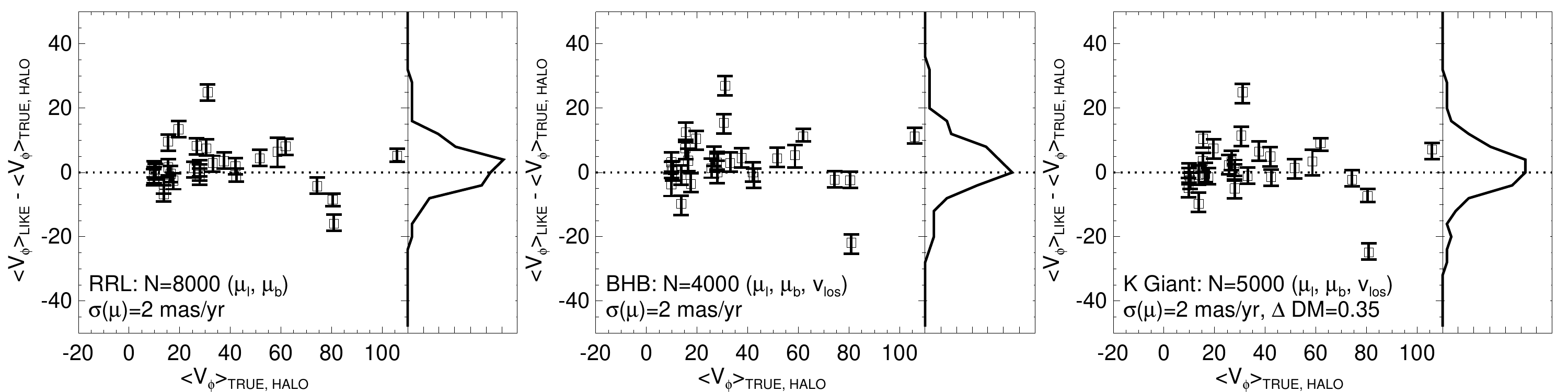}
    \caption{\small A comparison between the ``true'' rotation signal in the Auriga haloes and the value estimated by applying our likelihood procedure to mock observations.  Old halo stars ($T_{\rm form} > 10$ Gyr) in the simulations are identified in the range  $5 < r/\mathrm{kpc} < 50$, and $N \sim 4000-8000$ are randomly selected within the SDSS footprint with $|z| > 4$ kpc. The positions and velocities of the stars are converted into Galactic coordinates, and we apply a scatter of $2$ mas/yr when converting to proper motions. The left, middle and right panels show RRL, BHB and K giant ``like'' mocks. The RRL mocks have $N \sim 8000$ stars randomly selected and the line-of-sight velocity is presumed to be unknown. All three Galactic velocity components are used for the BHB and K giant mocks, but the sample size is smaller ($N \sim 4000-5000$). Finally, for the K giant mocks, we also apply a scatter of 0.35 dex to the distance moduli of the stars. The error bars show the 1$\sigma$ confidence derived from the likelihood analysis. The true rotation signal is typically recovered to within 5 km s$^{-1}$ (median $\sim 1$ km s$^{-1}$, $\sigma = 1.48 \times \mathrm{MAD} \sim 5$ km s$^{-1}$).  Note that the outliers with $|\langle V_\phi \rangle_{\rm LIKE} - \langle V_\phi \rangle_{\rm TRUE, HALO}|  > 20$ km s$^{-1}$ are likely due to substructure in the Auriga haloes. }
    \label{fig:mock}
\end{figure*}

\subsubsection{Tests with mock observations}
In Figure \ref{fig:sim_rot} we showed the ``true'' average rotation signal of the Auriga stellar haloes. This is computed for all halo stars within the SDSS footprint with $5 < r/\mathrm{kpc} < 50$ and $|z| > 4$ kpc directly from the simulations. Now, we generate mock observations from the simulated stellar haloes to see if we can recover this rotation signal using the likelihood method described in Section \ref{sec:like}. For the mock observations, we convert spherical coordinates ($r, \theta, \phi$) into Galactocentric coordinates ($D, \ell, b$), and place the ``observer'' at the position of the Sun $(x,y,z)=(-8.5, 0,0)$ kpc.  Old halo stars are identified ($T_{\rm form} > 10$ Gyr) in the coordinate ranges $5 < r/\mathrm{kpc} < 50$ and $|z| > 4$ kpc, and $N \sim 4000-8000$ are randomly selected within the SDSS footprint (see Fig. \ref{fig:sdss_rrl}). The tangential Galactic velocity components ($V_\ell, V_b$) are converted into proper motions, and we apply a scatter of 2 mas/yr, which is the typical observational uncertainty in the SDSS-\textit{Gaia} sample. After applying our modeling technique, we show the resulting best-fit $\langle V_\phi \rangle$ parameters in Fig. \ref{fig:mock}. The left, middle and right panels show RRL-, BHB- and K giant-like mocks. The RRL mocks have $N \sim 8000$ stars randomly selected and we marginalise over the line-of-sight velocity. All three Galactic velocity components are used for the BHB and K giant mocks, but the sample sizes are smaller ($N \sim 4000-5000$), and we apply a scatter of 0.35 dex to the distance moduli of the ``K giant'' stars. Note that we also use these mocks to ensure that we can safely ignore the small ($\sim 10\%$) distance uncertainties in the RRL and BHB populations. In Fig. \ref{fig:mock} we show the difference between the true and inferred $\langle V_\phi \rangle$ values as a function of the true rotation signal. The distribution of $\Delta \langle V_\phi \rangle = \langle V_\phi \rangle_{\rm LIKE} - \langle V_\phi \rangle_{\rm TRUE, HALO}$ is similar for all three mock tests, with median offset of $\sim 1$ km s$^{-1}$ and $\sigma=1.48 \times$ MAD of $\sim5$ km s$^{-1}$ (see right-hand inset)\footnote{Note that we attribute the outliers with large $\Delta \langle V_\phi \rangle$ to significant substructures in the Auriga haloes.}. Thus, even with observational proper motion errors of order the proper motions themselves, we are able to recover the average rotation signal of the stellar halo to $< 10$ km s$^{-1}$. Note that this level of scatter in the simulations is typically less than the \textit{systematic} uncertainty in the SDSS-\textit{Gaia} proper motion catalog of 0.1 mas/yr.

\section{Conclusions}
\label{sec:conc}

We have combined the exquisite astrometry from \textit{Gaia} DR1 and recalibrated astrometry of SDSS images taken some $\sim 10-15$ years earlier to provide a stable and robust catalog of proper motions. Using spectroscopically confirmed QSOs, we estimate typical proper motion uncertainties of $\sim 2$ mas/yr down to $r \sim 20$ mag, which are stable to variations in colour and magnitude. Furthermore, we estimate systematic errors to be of order 0.1 mas/yr, which is unrivaled by any other dataset of similar depth. We exploit this new SDSS-\textit{Gaia} proper motion catalogue to measure the net rotation of the Milky Way stellar halo using RRL, BHB and K giant halo tracers.
Our main conclusions are summarised as follows.

\begin{itemize}

\item We identify (RRL) halo stars that belong to the Sgr stream and compare the SDSS-\textit{Gaia} proper motions along the stream to the \cite{law10} model. In general, there is excellent agreement with the model predictions for the Sgr leading and trailing arms. Furthermore, previous proper motion measurements in the literature of the Sgr stream \citep{carlin12, koposov13, sohn15} agree very well with the new SDSS-\textit{Gaia} proper motions. These comparisons are a reassuring validation that these new proper motions can be used to probe the Milky Way halo.

\item We construct samples of RRL, BHB and K giant stars in the halo with measured proper motions, distances, and (for the spectroscopic samples) line-of-sight velocities. Using a likelihood procedure, we measure a weak prograde rotating stellar halo, with $\langle V_\phi \rangle \sim 5-25$ km s$^{-1}$. This weakly rotating signal is similar for all three halo samples, and varies little with Galactocentric radius out to 50 kpc. In addition, there is tentative evidence that the rotation signal correlates with metallicity, whereby metal-richer BHB and K giant stars exhibit slightly stronger prograde rotation. 

\item The state-of-the-art Auriga simulations are used to compare our results with the expectations from the $\Lambda$CDM model. The simulated stellar haloes tend to have a net prograde rotation with $0 \lesssim V_{\phi}/\mathrm{km s^{-1}} \lesssim 120$. However, when we compare with ``old'' ($T_{\rm form} > 10$ Gyr) halo stars in the simulations,  which are more akin to the old halo tracers like BHBs and RRL, the prograde signal is weaker and typically $V_{\phi} \lesssim 80$ km s$^{-1}$, in good agreement with the observations. Metal-rich(er) halo stars in the simulations are biased towards stronger prograde rotation than metal-poor(er) halo stars. It is likely that this correlation is, in part, due to contamination by disc stars and/or halo stars formed \textit{in situ}, which are more (kinematically) akin to a disc component. However, the rotation signal of the old halo stars, which are likely dominated by accreted stellar stars, shows only weak, if any, dependence on metallicity. Again, this is in line with the observations.

\item The weak prograde rotation of the Milky Way halo is in agreement with the simulations, but is still relatively low compared to the full Auriga suite of 30 haloes ($\sim$ 20th percentile). It is also worth remembering that the net spin of the halo disappears entirely if the circular velocity at the position of the Sun is set to the ``standard'' 220 km s$^{-1}$. Furthermore, the systematic uncertainty in the SDSS-\textit{Gaia} proper motions of $\sim 0.1$ mas/yr means that rotation signals $\lesssim 10$ km s$^{-1}$ are also consistent with zero. This mild, or zero, halo rotation suggests that above $z=4$ kpc, the Milky Way has (a) a minor, or non-existent, \textit{in situ} halo component and, (b) undergone a relatively quiescent merger history.

\item Finally, we use the simulated stellar haloes to quantify the systematic uncertainties in our modeling procedure. Using mock observations, we find that the rotation signals can typically be recovered to $< 10$ km s$^{-1}$. However, we do find that substructures in the halo can significantly bias the results. Indeed, in regions that the Sgr stream is prominent (e.g. $20 < r/\mathrm{kpc} < 30$) our measured rotation signal is increased by the Sgr members. 

\end{itemize}

 \section*{Acknowledgements}
We thank Carlos Frenk and Volker Springel for providing comments on an earlier version of this manuscript. We also thank the anonymous referee for providing valuable comments that improved the quality of our paper. 

A.D. is supported by a Royal Society University Research Fellowship. 
The research leading to these results has received funding from the
European Research Council under the European Union's Seventh Framework
Programme (FP/2007-2013) / ERC Grant Agreement n. 308024. V.B. and S.K. acknowledge financial support from the ERC. A.D. and S.K. also acknowledge the support from the STFC (grants ST/L00075X/1 and ST/N004493/1). RG acknowledges support by the DFG Research Centre SFB-881 ``The Milky Way System'' through project A1.

This work has made use of data from the European Space Agency (ESA)
mission {\it Gaia} (\url{http://www.cosmos.esa.int/gaia}), processed
by the {\it Gaia} Data Processing and Analysis Consortium (DPAC,
{\small
  \url{http://www.cosmos.esa.int/web/gaia/dpac/consortium}}). Funding
for the DPAC has been provided by national institutions, in particular
the institutions participating in the {\it Gaia} Multilateral
Agreement.

This work used the DiRAC Data Centric system at Durham University, operated by the Institute for Computational Cosmology on behalf of the STFC DiRAC HPC Facility (www.dirac.ac.uk). This equipment was funded by BIS National E-infrastructure capital grant ST/K00042X/1, STFC capital grants ST/H008519/1 and ST/K00087X/1, STFC DiRAC Operations grant ST/K003267/1 and Durham University. DiRAC is part of the National E-Infrastructure.

\bibliographystyle{mnras}
\bibliography{mybib}

\appendix

\section{QSO Proper Motions}
\label{sec:appendix}

\begin{figure*}
    \centering
    \includegraphics[width=16cm, height=10.67cm]{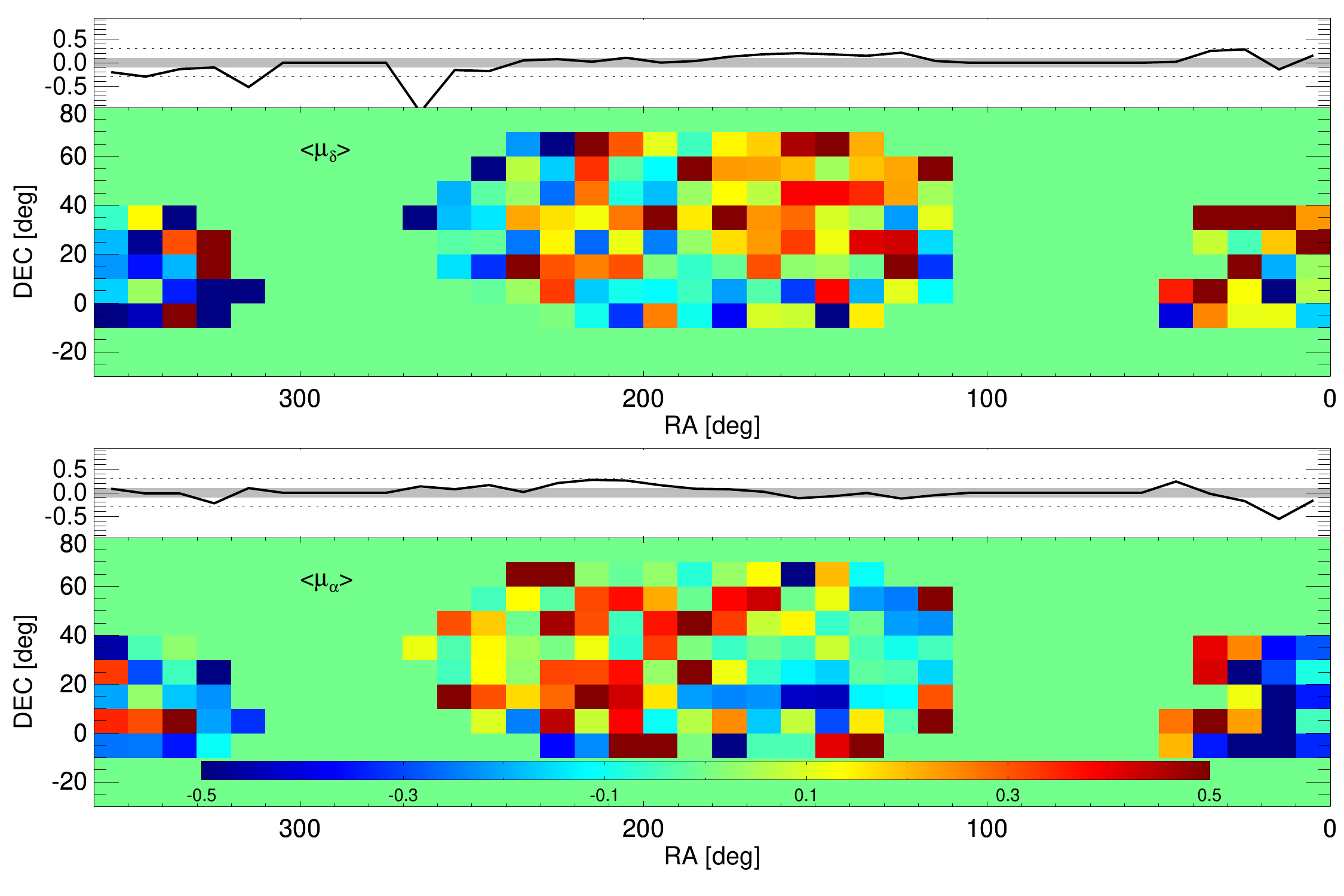}
    \caption[]{\small The median QSO proper motions in bins of RA and DEC. The $\mu_\delta$ and $\mu_\alpha$ components are shown in the top and bottom panels, respectively.  The inset panels show the median proper motions in bins of RA. The grey filled region indicates systematic offsets of $\pm 0.1$ mas/yr, and the dotted lines show deviations of $\pm 0.3$ mas/yr. We find no significant trends with systematics and position on the sky.}
    \label{fig:qso_radec}
\end{figure*}

In Fig. \ref{fig:qso_radec} we explore how the median QSO proper motions vary with position on the sky. We find that the systematics on the sky are at the level of 0.1-0.2 mas/yr with maximal (mostly non-systematic) deviations of 0.5 mas/yr. This is in stark contrast to what \cite{tian17} found for their \textit{Gaia}-PS1-SDSS proper motion catalog, where the QSO proper motions have systematic patterns with amplitudes of 2 mas/yr. \cite{tian17} suggest that these large variations could be due to differential chromatic refraction (DCR) induced motions in the QSOs. Although QSOs are appealing objects to test for proper motion uncertainties and systematics, the possibility of DCR affects is worrisome. However, for discernible DCR affects we would expect strong correlations with airmass and a QSO redshift dependence that does not average to zero (see Figure 3 of \citealt{kaczmarczik09}). By comparison with Figure 11 in \cite{leistedt13} we find little correlation between the QSO proper motions with airmass. Furthermore, we showed in Fig. \ref{fig:qso_mag_col} that there is little dependence on the QSO proper motion distributions with $g-r$ colour (and therefore redshift). We therefore conclude that DCR related affects in our proper motion catalog are minimal, and we can safely use QSOs to quantify our statistical and systematic proper motion uncertainties.

\bsp
\label{lastpage}
\end{document}